 \documentclass[final,3p,times,twocolumn]{elsarticle}

\usepackage{amssymb}

\usepackage{graphicx}
\usepackage[cmex10]{amsmath}
\usepackage[hyphens]{url}
\usepackage{color}
\usepackage{algorithmic}
\usepackage{algorithm}
\usepackage{subfigure}
\usepackage{microtype}
\usepackage{amsmath}
\usepackage{url}
\usepackage{subfigure}

\usepackage{amsbsy}
\usepackage{float}

\newcommand{\mr}[1]{\mathrm{#1}}

\long\def\comment#{}

\journal{Journal of Computational Physics}

\begin{document}

\begin{frontmatter}



\title{Parallel Tempering Simulation of the three-dimensional Edwards-Anderson Model  with  Compact Asynchronous Multispin Coding on GPU}


\author[a,b]    {Ye Fang}
\author[a,c]      {Sheng Feng}
\author[a,c]      {Ka-Ming Tam}
\author[a,d]      {Zhifeng Yun}
\author[a,c]    {Juana Moreno}
\author[a,b]    {J.~Ramanujam}
\author[a,c]    {Mark Jarrell}

\address[a]{Center for Computation and Technology, Louisiana State University, Baton Rouge, LA 70803, USA}
\address[b]{School of Electrical Engineering and Computer Science, Louisiana State University, Baton Rouge, LA 70803, USA}
\address[c]{Department of Physics and Astronomy, Louisiana State University, Baton Rouge, LA 70803, USA}
\address[d]{Center for Advanced Computing and Data Systems, University of Houston, Houston, TX 77204, USA}

\begin{abstract}

Monte Carlo simulations of the Ising model play an important role in
the field of computational statistical physics, and they have revealed
many properties of the model over the past few decades. However, the
effect of frustration due to random disorder, in particular the
possible spin glass phase, remains a crucial but poorly understood
problem. One of the obstacles in the Monte Carlo simulation of random
frustrated systems is their long relaxation time making an efficient
parallel implementation on state-of-the-art computation platforms
highly desirable. The Graphics Processing Unit (GPU) is such a
platform that provides an opportunity to significantly enhance the
computational performance and thus gain new insight into this
problem. In this paper, we present optimization and tuning
approaches for the CUDA implementation of the spin glass simulation on
GPUs. We discuss the integration of various design
alternatives, such as GPU kernel construction with minimal
communication, memory tiling, and look-up tables. We present a 
binary data format, Compact Asynchronous Multispin Coding (CAMSC),
which provides an additional $28.4\%$ speedup compared with the
traditionally used Asynchronous Multispin Coding (AMSC). Our overall
design sustains a performance of 33.5 picoseconds per spin flip attempt
for simulating the three-dimensional Edwards-Anderson model with parallel tempering,
which significantly improves the performance over existing GPU implementations.


\end{abstract}


\end{frontmatter}



\section{Introduction}
%
%
%
%

Stochastic or Monte Carlo (MC) simulation is one of the most important methods in 
the study of complex interacting systems. However, even with the huge success 
of Monte Carlo methods, many systems remain very difficult to simulate. 
The main obstacle very often is the long required simulation time, while 
the memory demands are quite modest. A prominent example is the Edwards-Anderson (EA) 
model, where the inherent randomness and frustration lead to very long relaxation 
times. Although the EA model has been intensively simulated over the 
past few decades, including implementations using gate-level reconfigurable processors 
\cite{Monaghan-1993} and some dedicated computers designed specifically for solving this model, 
\cite{Ogielski-Morgenstern-1985,Ogielski-1985,Cruz-2001,Condon-Ogielski-1985,Taiji-Ito-Suzuki-1988} 
many aspects are still far from completely understood. Some prominent 
topics, such as the nature of the spin glass phase below the upper critical 
dimension, remain highly debated issues.~\cite{Jorg-Katzgraber-Krzakala-2008,Moore-2005,Young-Katzgraber-2004,Temesvari-2008,Katzgraber-2008,Sasaki-etal-2008,Sasaki-etal-2007,Larson-etal-2013,Banos2012,Katzgraber-2012,Katzgraber-Larson-Young-2009,Leuzzi-2009}

The Graphics Processing Unit (GPU) provides an opportunity to 
significantly improve the computational performance of Monte Carlo simulations 
of classical systems. Massive parallelism and acceleration can be achieved by 
implementing these algorithms on GPUs. In the past few years some GPU accelerated 
simple spin models have been proposed, including the two-dimensional Ising model 
by \citet{CSTN-093} and \citet{2010CoPhC.181.1549B}, and the Ising model in the cubic 
and network lattices by \citet{Preis:2009:GAM:1537305.1537344}.
\citet{doi:10.1142/S0129183112400025,Weigel:2012:PPS:2151219.2151631}
studied the Ising and the Heisenberg models 
in both two- and three-dimensional lattices. These implementations focus 
predominately on unfrustrated systems with large lattice sizes. In this study, we mainly 
focus in the simulation of a random frustrated Ising system in equilibrium. 
Due to its slow relaxation rate, a large number of Monte Carlo 
steps are required, at the same time the system sizes that can be simulated are 
relatively small, in most cases limited to only a few thousands sites. Precisely 
because of these characteristics, Monte Carlo simulations of random frustrated 
systems are a good match for the GPU computing architecture.

Our implementation targets cluster computers with NVIDIA Fermi GPUs.
Using C/CUDA we control and tune details of the program.
We expose the inherent parallelism of the algorithm
to the GPU accelerator, including parallel computation on multiple sites, multiple temperature 
replicas and multiple disorder realizations. The memory requirements are efficiently handled through
memory tiling. In addition, the computation is simplified and
vectorized using table look-ups and the Compact Asynchronous
Multispin Coding (CAMSC). We also substitute all floating point arithmetic
with integer or bit string computations while preserve the same
precision. Combining various tuning techniques, we achieve an average
spin flip time of 33.5 picoseconds. This is the fastest GPU
implementation for the random frustrated Ising system on a $16^3$
cubic lattice, and is comparable to that obtained with a field
programmable gate array (FPGA) hardware \cite{2012arXiv1204.4134J} for
small to intermediate system sizes. We note that a very recent preprint reported a faster speed in
a new FPGA system \cite{Janus2-2013}. 

The paper is organized as follows. In Section 2, we discuss the algorithm. In section 3, 
we present an outline of the code framework. The implementation 
and optimization methods are described in Section 4. Section 5 shows the experimental  
results. Conclusions and future directions are described in Section 6.

\section{Theoretical Background}

\subsection{Spin Glass}

The discovery of a plethora of unusual magnetic behaviors in disordered materials 
initiated the field of glassy systems.\cite{Binder-Young1986} Spin glasses are 
beyond the conventional description of long range magnetic ordering, e.g., 
ferromagnetic ordering. Some of their features, including their frequency-dependent 
susceptibilities and the discrepancy between zero-field and field cooling measurements, 
suggest that spin glasses have very slow dynamics. Notwithstanding most experimental spin 
glass systems, which exhibit glassy behavior, randomness and frustration 
seem to share some common properties. In real materials, 
dilution introduces randomness and directional or distance-dependent couplings, 
such as dipolar interactions in insulating systems and the Ruderman-Kittel-Kasuya-Yoshida 
coupling in metallic systems, introduce frustration. 

The simplest model that captures the consequences of disorder is an Ising model 
with quenched randomly disordered couplings. This model was first proposed by Edwards 
and Anderson. \cite{Edwards-Anderson1975} The mean field solution of the EA 
model for infinite dimensions was first attempted by Sherrington and Kirkpatrick. 
\cite{Sherrington-Kirkpatrick1978} However, the replica symmetric mean field solution was found to be 
unstable below the Almeida-Thouless line, \cite{Almedia-Thouless1978,Bray-Moore-1978}
a line in the temperature-field plane below which replica symmetry is broken. The difficulty 
of obtaining a stable solution was solved by Parisi with his replica symmetry breaking 
ansatz. \cite{Parisi-1979,Mezard-etal-1984,Parisi-1980a,Parisi-1980b,Parisi-1980c,Parisi-dirac-medal-2002} 
Although the mean field solution has been proven to provide the exact free energy for the spin glass phase 
in infinite dimensions, \cite{Talagrand-2006,Guerra-2003} the spin glass 
physics in finite dimensions, which presumably is more relevant to experiments, is 
still not fully understood. Indeed, it had long been debated whether a spin glass 
phase at finite temperatures exists in three dimensions.

The EA model may be deceptively simple. Since it is a classical spin 
model, one may think that its numerical study can be simply carried out by Monte 
Carlo methods on conventional hardware. One of the defining signatures of 
spin glass systems is their long relaxation time. For sufficiently low temperatures, the 
system becomes very sluggish and equilibration is prohibitively difficult  
even for modest systems sizes.  Moreover, it has been shown
that finding the ground state of the three dimensional EA model is
an NP-hard problem. \cite{Barahona-1982} Until recently, there has been no 
consensus on whether there is a finite spin glass critical temperature in the three 
dimensional EA model.

The breakthrough in the numerical study of spin glass systems came with the 
introduction of the parallel tempering method. It allowed the study of larger 
systems at lower temperatures than the simple single spin flip 
method. \cite{Swendsen-Wang-1986,Hukushima-Nemoto1996,Marinari-Parisi1992} Combined with 
improved schemes for finite size scaling, it is now widely believed that the thermodynamic 
finite-temperature spin glass phase does exist in the three dimensional EA 
model~\cite{Ballesteros2000}. As the upper critical dimension of the 
model is six \cite{Harris-Lubensky-Chen-1976,Tasaki-1989,Green-Moore-Bray-1983}, a
prominent remaining question is the nature of the 
spin glass phase below the upper critical dimension~\cite{Young-Katzgraber2004}. 
In particular, if the spin glass can still be described by the replica symmetry 
breaking scenario, there should be an Almeida-Thouless line below the upper 
critical dimension. A possible test of whether the Almeida-Thouless 
line exists is to determine whether a spin glass phase exists under an external 
magnetic field. Correlation length scaling analysis seems to suggest the 
absence of the spin glass phase in cubic lattices when a finite external field is applied.\cite{Young-Katzgraber-2004} 
On the other hand, a recent study in four-dimensional lattices suggests that by using a different quantity for the 
finite size scaling analysis, a spin glass phase can be revealed. \cite{Banos2012} 
Given the relevance of spin glasses and the on-going controversy on the nature 
of the spin glass phase below the upper critical dimension, it is desirable to 
implement an efficient parallel tempering Monte Carlo algorithm using 
graphics processing units to accelerate the simulations. In this work we show that 
using the multispin coding method, \cite{Zorn-Herrmann-Rebbi-1981} an efficient Monte Carlo algorithm can be implemented 
on the GPU.

\subsection{Edwards-Anderson Model}

We consider the EA Model \cite{Edwards-Anderson1975} on a 
simple cubic lattice. Spins on each lattice site have two states 
$S_i=+1$ or $-1$. The couplings $J_{ij}$ are between nearest neighbors. In this 
study, we focus on a distribution of the couplings which is bimodal with a mean 
value of zero. That is, there are equal numbers of anti-ferromagnetic and ferromagnetic 
couplings. The effect of the distribution is certainly a non-trivial problem. 
We choose to focus on the bimodal distribution  because it is best suited for multispin coding. 
In addition, a constant external field, $h$, is included in our implementation. 
The Hamiltonian is given by 

\begin{equation}
H = - \sum_{i,j} S_i J_{ij} S_j + h \sum_{i} S_i.
\end{equation}

\subsection{Single Spin Flip Metropolis Algorithm}

We implement the Metropolis algorithm as our sampling method. The spins are 
visited and tested for flipping according to the probability $P = \mr{exp}(-\beta\Delta E)$,
where $\beta$ is the temperature and $\Delta E$ is the energy change associated 
with the proposed spin flip. As the algorithm satisfies detailed balance, the sampling 
will generate a distribution according to the partition function provided that the 
simulation is performed long enough.  This type of Monte Carlo simulation is
called a Markov process, because the evolution of the state only depends on the state 
at the current step, and not on its history. 

\subsection{Parallel Tempering}

For the simulation of glasses, the local single spin 
update algorithm is very slow in thermalizing the system. This problem is particularly 
severe when the temperature is close to the critical temperature for the second order 
transition. For certain spin glass models where random dilution is sufficiently 
large, some form of cluster algorithm can improve the rate of thermalization. Unfortunately, 
there is no efficient cluster methods for general spin glass systems. The possible exceptions are 
some random diluted systems or systems in low dimension \cite{Houdayer-2001,Liang-1992,Jorg-2005}. Various other methods have been proposed 
in the past to improve the rate of thermalization including the umbrella sampling, 
the multi-canonical method, and rejection-free methods. It is now widely accepted that 
the parallel tempering method is one of the most efficient algorithms for improving 
the thermalization rate of general spin glass systems.

\begin{figure}[ht]
  \centering
  \includegraphics[width=0.4\textwidth] {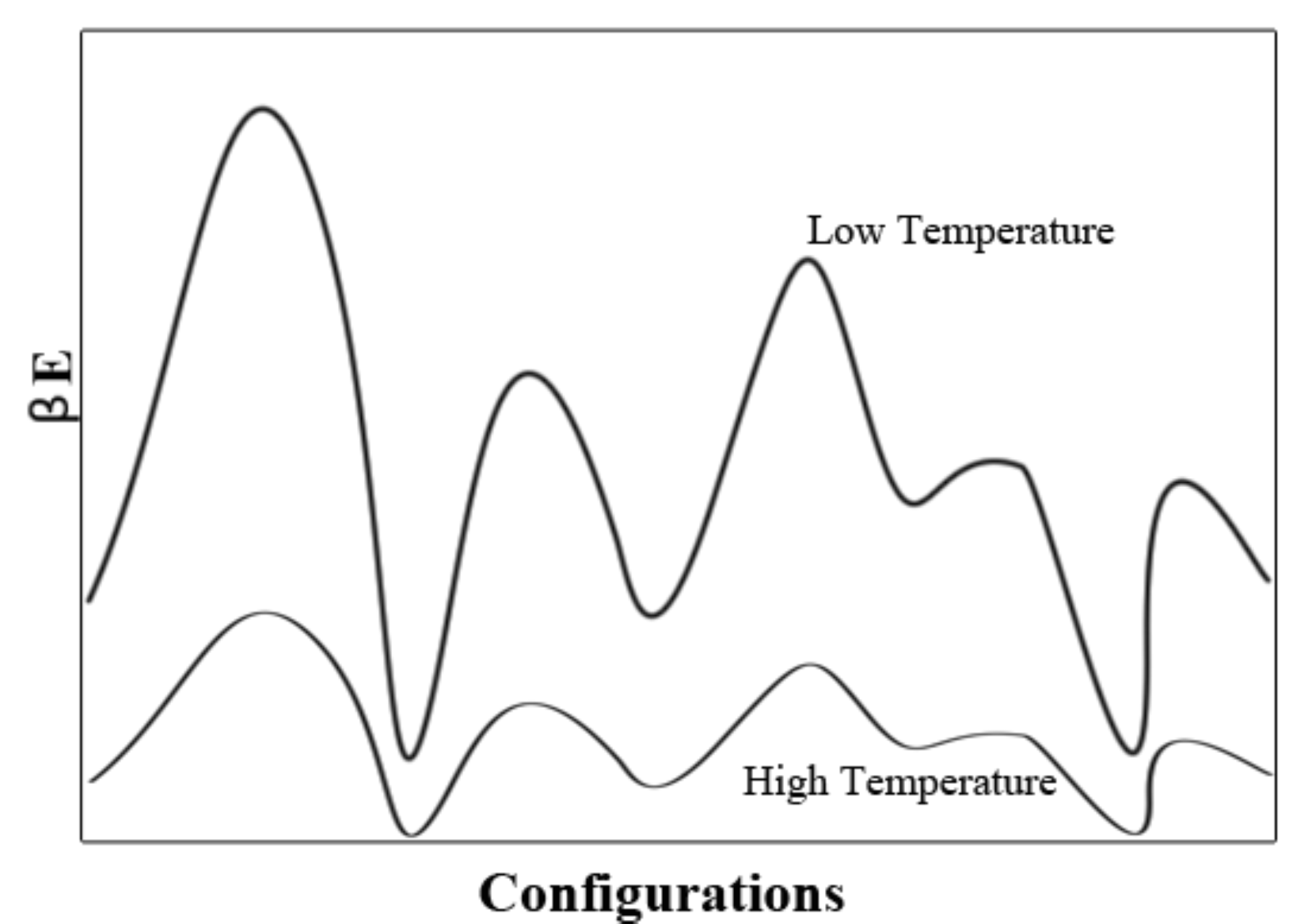}
  \caption{Schematic diagram of the free energy landscape.  At high temperatures 
(small $\beta$) the barriers between configurations are reduced allowing the 
system to search through configurations more efficiently.}
\label{fig-pt}
\end{figure}

Parallel tempering uses several samples of the system within a range of 
temperatures (Figure \ref{fig-pt}). The low temperature sample is more difficult 
to thermalize due to the larger barriers between low energy 
configurations~\cite{Marinari-Parisi1992,Hukushima-Nemoto1996}.
However, as the probability to swap the configuration between the high and the low temperature samples 
increases, the chance of the system to escape from a local minima in the low temperature sample also 
increases. The efficiency of such a parallel 
tempering move can be measured by the time it takes for a sample to perform a 
round trip along the temperature axis, that is from the lowest to the highest 
temperature and back to the lowest temperature. This largely depends on 
the system being simulated. Fine tuning the range of temperatures and the 
spacing between them is crucial to optimize the performance. Some 
recent proposals have been tested on the non-disorder Ising model \cite{PhysRevLett.101.130603,1742-5468-2006-03-P03018,jcp/124/17/10.1063/1.2186639}.
Models with explicit disorder such as the EA lack an efficient 
general method. For a practical GPU implementation, one also 
needs to consider the effect of the number of replicas on the performance.

\section{The Framework}

\label{section_framework}

The GPU implementation is discussed in this and the following sections. 
In our replica exchange spin glass simulation we exploit three levels of parallelism:
\begin{enumerate}
\item Several tens of thousands, or more,  of independent disorder realizations are required to 
obtain good statistics. 
\item For each disorder realization, usually a few tens of systems at different temperatures   
are needed to study the physics, such as the possibility of a critical point. We 
denote these systems as temperature replicas.   In the parallel 
tempering simulation, different temperature replicas  communicate with each other only during 
the parallel tempering swap; these swaps are performed after every few Metropolis single spin sweeps 
of the lattice. 
\item We are mainly interested in systems on bipartite lattices. These are lattices that can 
be divided in two sub-lattices (A and B) with same sub-lattice spins 
do not directly coupling with each other. As a result, the update of the A sublattice 
is independent of the B sublattice. 
\end{enumerate}
These three levels of inherent parallelism allows an efficient GPU
implementation. In this section we focus on the main
structure of the code, which consists of three parts: (i) distributing
the spin updates into different GPU threads; (ii) distributing 
different disorder realizations into different GPU
blocks; and (iii) integrating and vectorizing the bit computations
of many temperature replicas

\subsection{Map Lattice Sites to GPU Threads}

\label{section_framework_threads}

\begin{figure}[ht]
  \centering
  \includegraphics[width=0.35\textwidth] {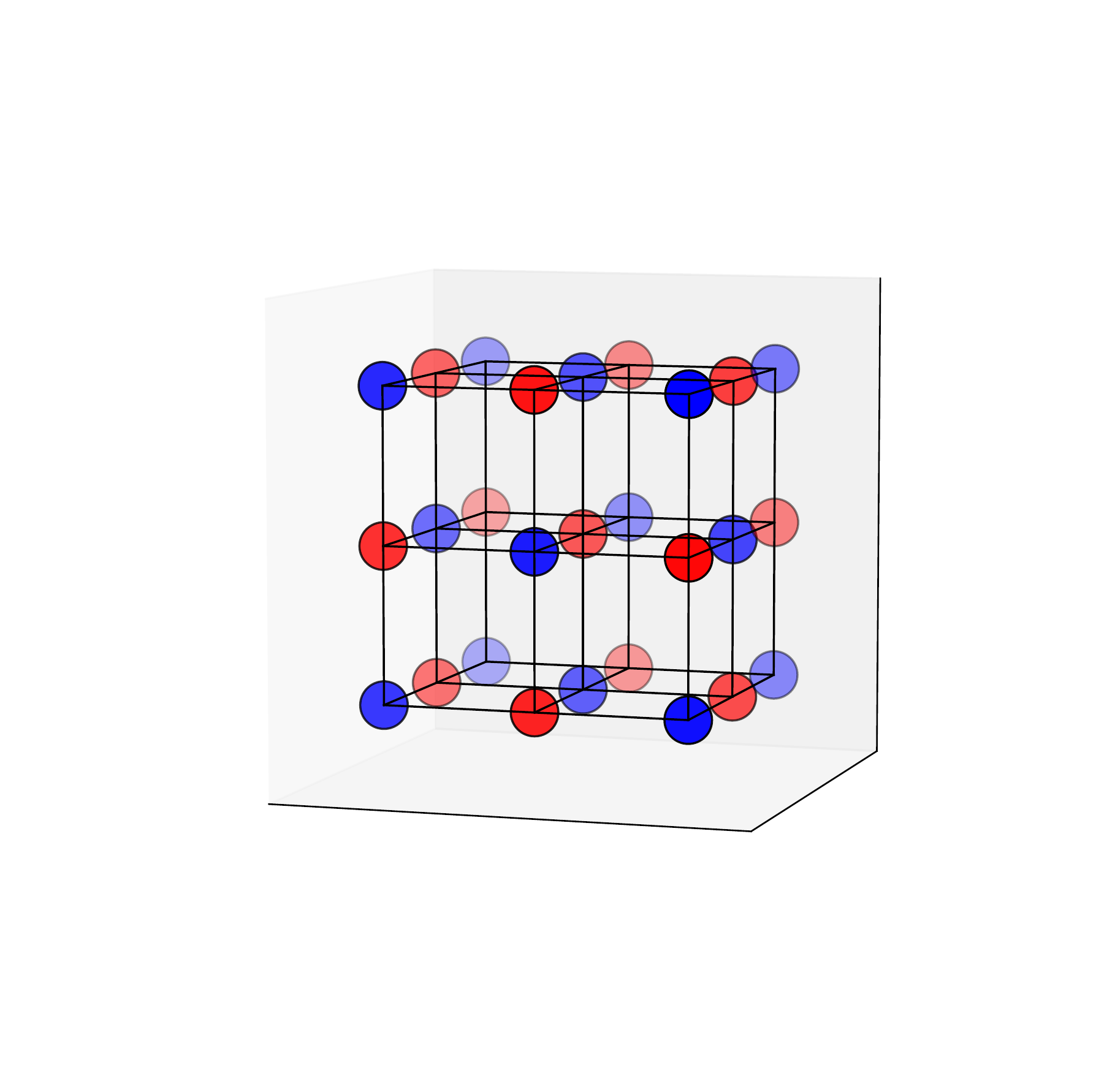} 
  \caption{A demonstration of the 3D checkerboard decomposition. 
The blue and red sites are on different sub-lattices. 
Since the sites in a sub-lattice never directly interact with each other, 
it is permissible to update different sites in parallel. 
}

\label{fig_checkerboard}
  \end{figure}

The spin lattice is represented by a three dimensional primitive cubic system. To update the sites in 
the lattice, we follow the common practice of employing a checkerboard decomposition that 
splits the sites into two sub-lattices shown in blue and red in Figure \ref{fig_checkerboard} . 
Since a blue site is surrounded by red sites and never directly interacts with other 
blue sites and vice-versa, it is permissible to update each sub-lattice in parallel.  
We construct two consecutive stages concentrating 
independently on each of the sub-lattices for parallel computation. The combination of 
the two stages delivers a lattice sweep of Monte Carlo updates. The lattice is assigned 
to a GPU thread block, and sites are split across the threads. Details about the 
lattice site to thread mapping will be discussed in Section \ref{section_memory} 
where we discuss memory optimizations. The total available thread-level parallelism 
is half of the total lattice sites, and specifically, falls into the range between 
$8^3 / 2 = 256$ to $16^3 / 2 = 2048$ since our simulation targets lattices 
between $8^3$ to $16^3$ sites.

\subsection{Map Temperatures Replicas to Bits} 

\begin{figure}[ht]
  \centering
  \includegraphics[width=0.35\textwidth] {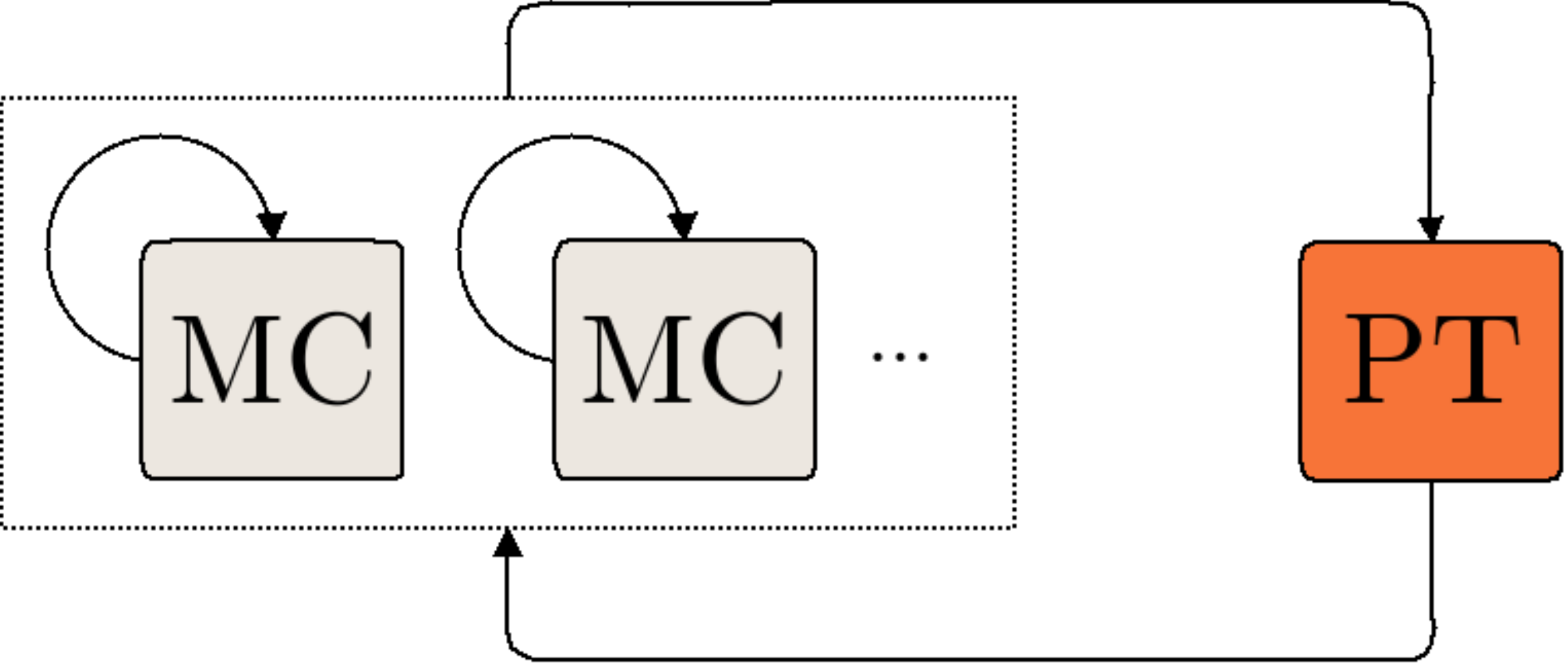}
  \caption{Many temperature replicas can be simulated simultaneously, each 
using an independent Monte Carlo process. These replicas may be exchanged
after a configurable steps of updates. 
A single GPU thread block is responsible for updating all the Monte Carlo processes
and manipulating the parallel tempering exchange.
}
 \label{fig_bits}
  \end{figure}

The parallel tempering technique facilitates the systems to achieve 
equilibrium. We choose the temperature as the tempering parameter and generate systems 
with the same couplings but different temperatures, called temperature replicas. The 
temperature replicas are uncorrelated during the spin-flip process and can therefore
be updated in parallel. However, they communicate and 
swap temperatures (Figure \ref{fig_bits}) after a few Monte Carlo sweeps. To better 
utilize the parallelism of multiple temperature replicas and minimize the communication 
overhead we have developed the Compact Asynchronous Multispin Coding (CAMSC), where 
spins from different temperature replicas at the same position are encoded into 
an integer. This leads to sub-word vectorization and a significant reduction of 
memory transactions. Details of our multispin coding procedure can be found in 
Section \ref{section_msc}. The number of temperature replicas depends on the 
system size and the temperature range. In our simulation we used 24 replicas 
for smaller systems, and 56 temperatures for bigger systems (for example, 
$10^3$ and $12^3$).

\subsection{Map Realizations to GPU Blocks}

\begin{figure*}[ht]
  \centering
  \includegraphics[width=0.95\textwidth] {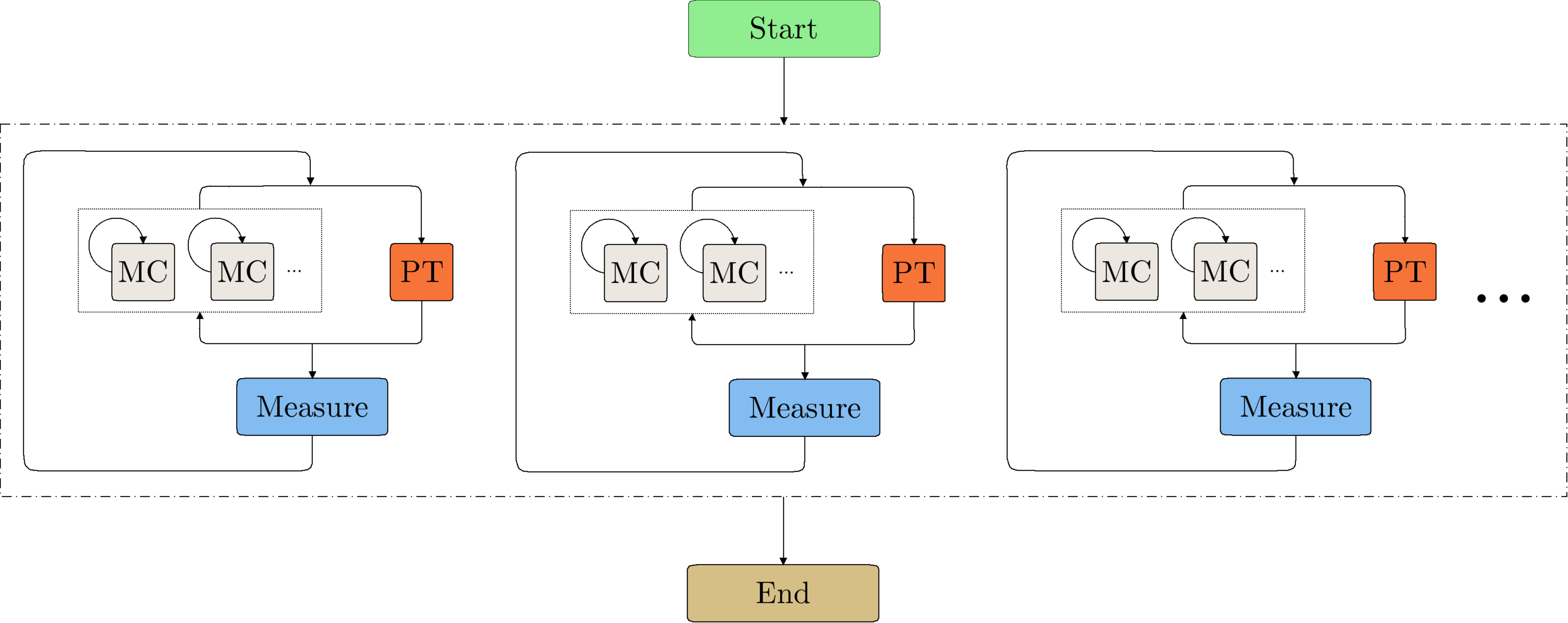}
  \caption{The outline of the simulation application. Disorder realizations are completely 
independent and can run simultaneously. Each realization contains a unique 
Monte Carlo parallel tempering process as depicted in Fig.~\ref{fig_bits}, and is 
assigned to a GPU thread block. This task level parallelism yields sufficient number of  
thread blocks and can fully occupy a parallel computer system.
}
  \label{fig_tasks}
\end{figure*}

Spin glass simulations usually require a larger number of disorder realizations ($10^4$ or more) for reliable 
disorder averaging.  A realization including all temperature replicas has been designated to 
a thread block.  We launch numerous thread blocks across multiple GPUs of multiple hosts 
until we get the sufficient number of realizations for disorder averaging (Figure \ref{fig_tasks}). To distribute 
these jobs across multiple nodes, we employ a Pthreads/MPI wrapper for the job distribution. 

\subsection{Discussion}

Some parallel processes are sequentialized for better memory
locality. For example, although the temperature replicas could be
fully parallelized as individual tasks or a lattice may be partitioned
across multiple thread blocks, we avoid these forms of parallelism.
The remaining parallelism is rich enough (with $10^4$ or more 
thread blocks) to fully occupy the cluster. 


To evaluate 
the performance, we employ a performance metric of
average time (in picoseconds) per proposed spin flip for a single GPU card:
\begin{equation}
  \label{eq:tsf}
  t=T_\mathrm{total} / N_\mathrm{MCS} / \left( N_\mathrm{spins} \times N_T \times N_\mathrm{samples} \right),
\end{equation}
where $T_\mathrm{total}$ is the total wall time of a simulation;
$N_\mathrm{MCS}$ is the number of Monte Carlo sweeps;
$N_\mathrm{spins}$ is the number of spins within a lattice;
$N_T$ is the number of temperature replicas;
$N_\mathrm{samples}$ is the total number of disorder realizations on one GPU card.
We develop and benchmark the code on a NVIDIA GeForce GTX 580 GPU. Detailed
platform configurations can be found in Section \ref{section_exp}.

\section{Implementation}

\label{section_implementation}

We discuss implementation details in this section, including the
construction of the GPU kernel, memory optimization, and various
techniques used to simplify the computation.

\subsection{Kernel Organization Optimization}

\label{section_korg}

Our simulation starts with the Pthreads/MPI job dispatcher that 
forks many CPU processes across the cluster computer system. Each CPU
process is responsible for initiating a
lattice realization, which is offloaded to its attached GPU 
for simulation until the spin variables or thermal averaged results are 
retrieved from the GPU back to the CPU for analysis. 

\begin{figure}[ht]
  \centering
  \includegraphics[width=0.35\textwidth] {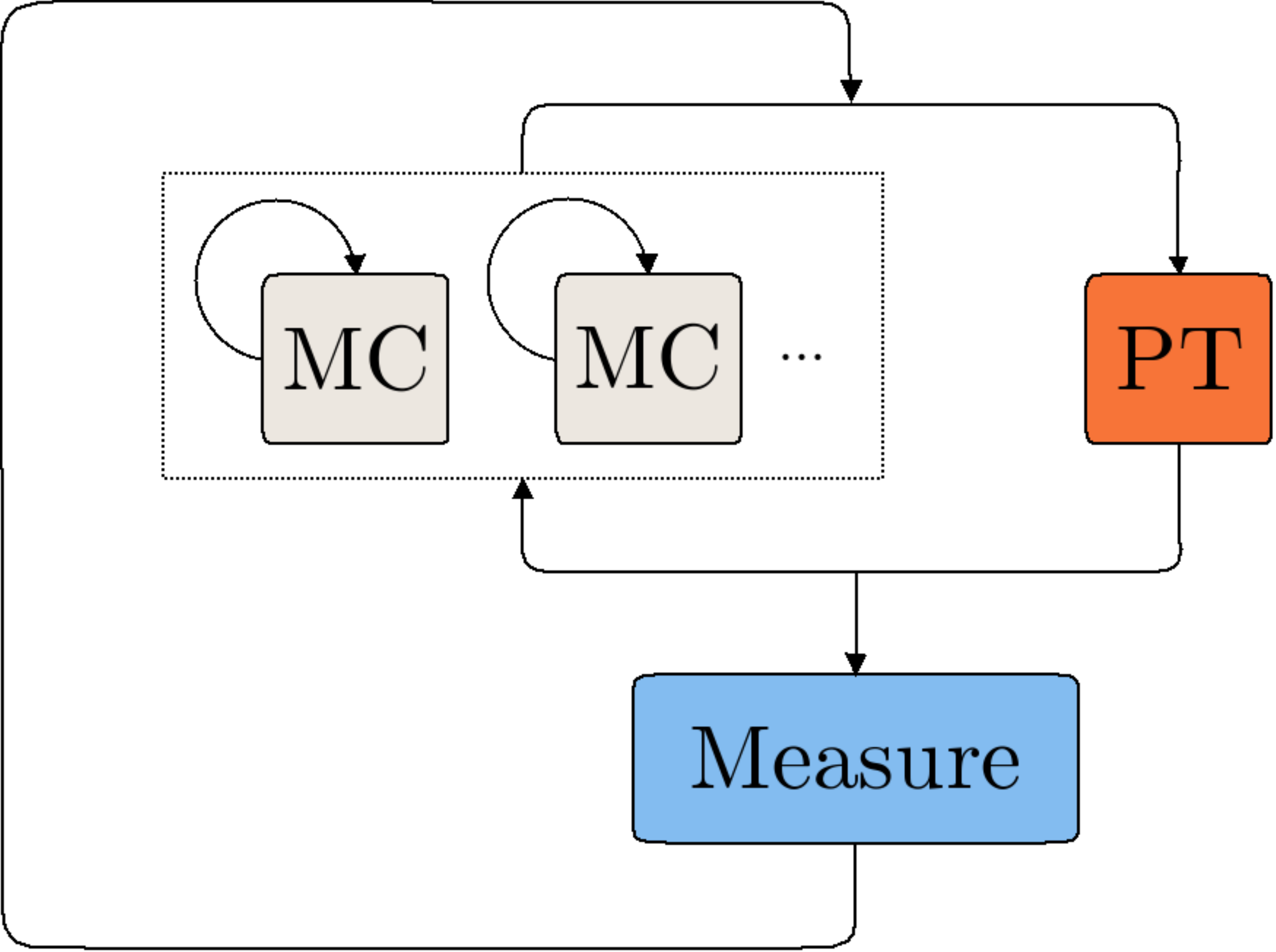}
  \caption{Three major components of the GPU program. 
One kernel calls Monte Carlo and parallel tempering, implemented as 
two device functions. Measurement is implemented as a separate GPU kernel.} 
  \label{fig_korg1}
  \end{figure}

The GPU workload has three major components (Figure \ref{fig_korg1}):
\begin{enumerate}
\item {\bf Metropolis moves}: The Metropolis steps for the single spin
  update for each temperature replica. This is done by calculating
  the local energy change and then comparing the acceptance ratio to a
  uniformly distributed random number.
\item {\bf Parallel tempering moves}: Parallel tempering swaps are
  performed after a few complete Monte Carlo sweeps of the
  lattice. This step requires the calculation of total energy for all
  temperature replicas; we use this to evaluate the acceptance ratio
  of parallel tempering swaps.
\item {\bf Measurements}: The spin configurations are dumped to the
  GPU global memory periodically to provide data for the
  measurements. In practice, we perform one measurement for every few
  thousands Metropolis sweeps.
\end{enumerate}

The measurement code has little overlap with the Monte Carlo and
parallel tempering codes, and it is called much less frequently. We
implement this part of the code as an separate GPU kernel.

\begin{figure}[ht]
  \includegraphics[width=0.5\textwidth] {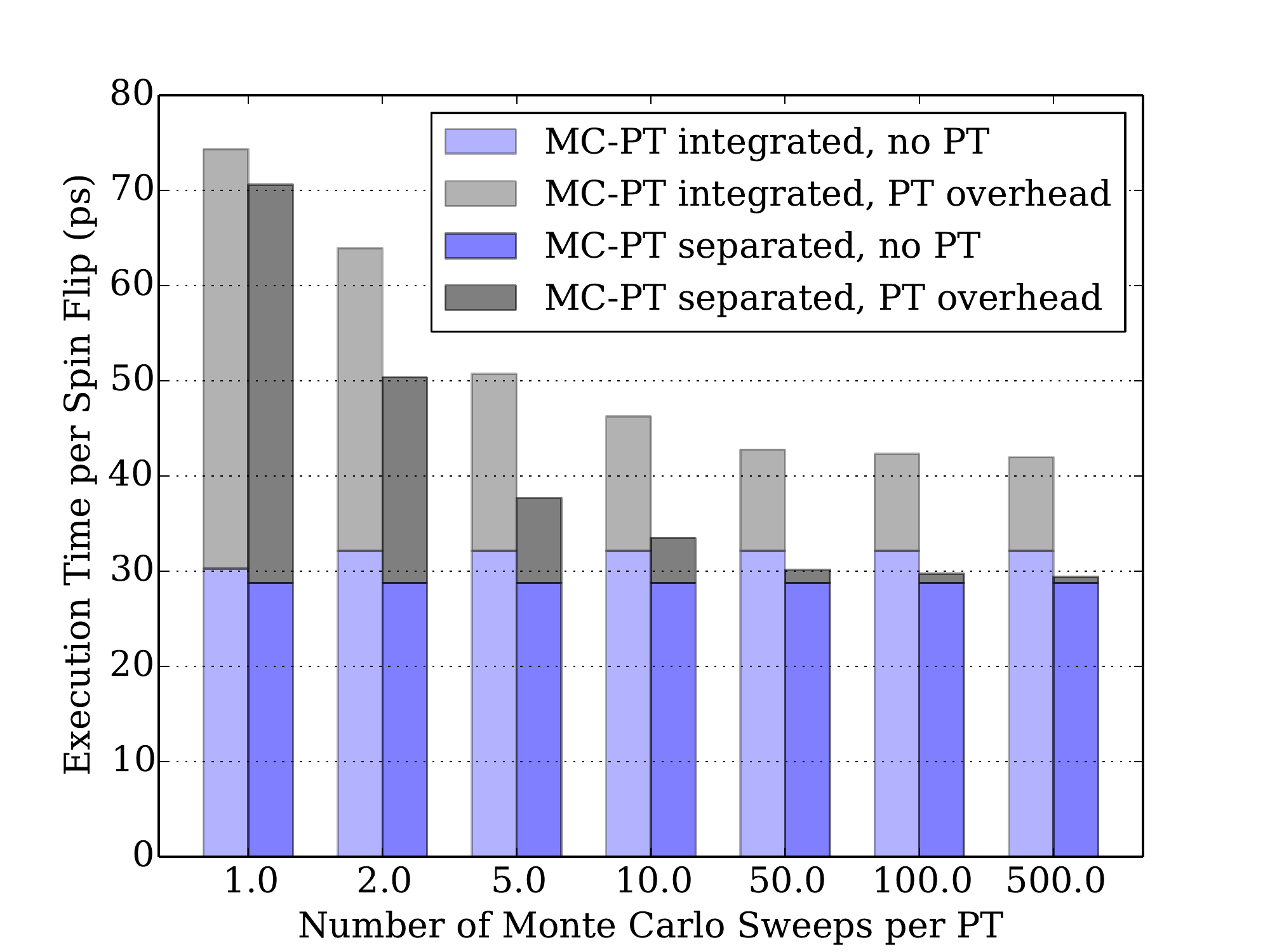}
  \caption{A comparison of the performance of the {\bf MC-PT integrated} and the {\bf MC-PT separated} schemes
with different numbers of Monte Carlo sweeps between an attempted parallel tempering swap.
The test is conducted with a $16^3$ cubic lattice, shared memory probability table of integers, CURAND, 
and CAMSC. }
  \label{fig_korg}
  \end{figure}

Both Monte Carlo and parallel tempering functions compute spin local
energies. Parallel tempering requires additional steps to sum the
local energies. Since an efficient implementation of sum (a form
of reduction)
consumes a considerable amount of shared memory, it may be efficient
to separate the parallel tempering as a dedicated GPU function
apart from the Monte Carlo. We denote this scheme {\bf MC-PT
  separated}. Alternatively, the {\bf MC-PT integrated} scheme
combines both the Monte Carlo and parallel tempering in a single GPU
kernel. Benchmarks (Figure \ref{fig_korg}) show that the 
{\bf MC-PT separated} scheme always performs better regardless of the frequency of
parallel tempering. However, we find that roughly 10 full Monte Carlo
sweeps of the lattice between parallel tempering attempts is a
speed/effectiveness sweet point.

\subsection{Memory Optimization}

\label{section_memory}

\begin{figure}[ht]
  \centering
  \includegraphics[width=0.25\textwidth] {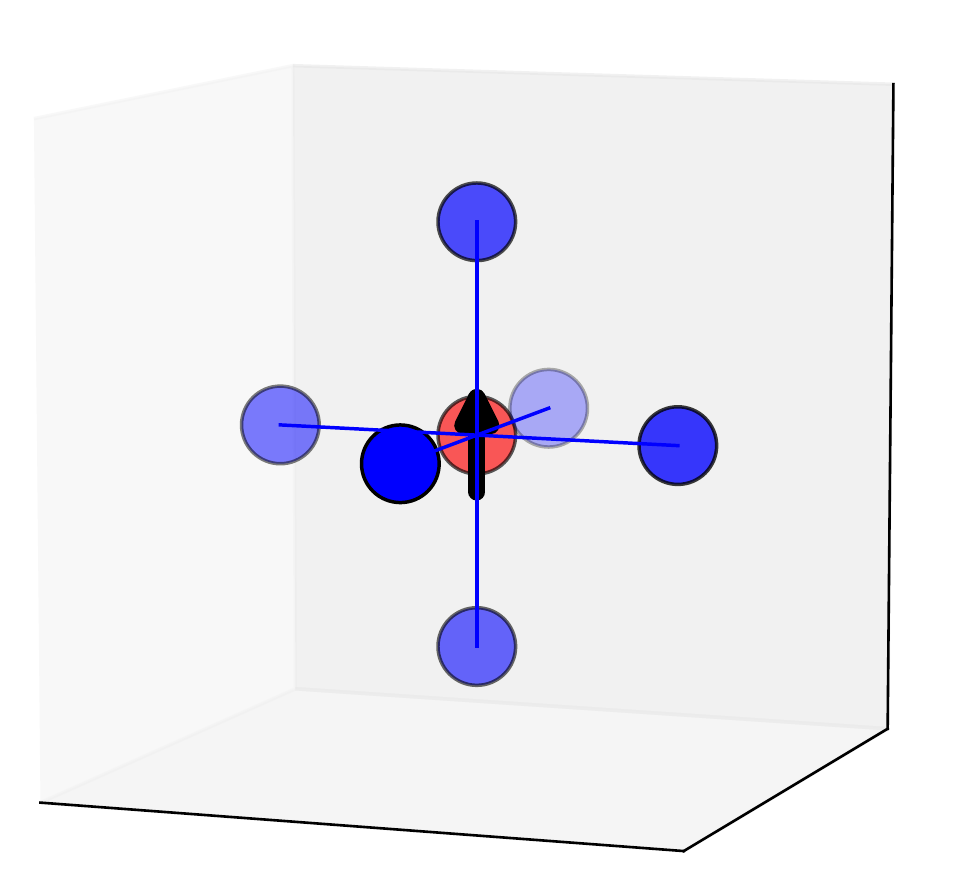}
  \caption{The memory access pattern for a single spin update where in
    addition to the state of the local spin, we also need the states
    of its 6 neighbors. Periodic boundary conditions are used.}
\label{fig_stencil}
\end{figure}

Each spin interacts with its six nearest neighbors (Figure
\ref{fig_stencil}) as a seven-point 3D stencil
\cite{Nguyen:2010:BOS:1884643.1884658, Datta:2008:SCO:1413370.1413375} with periodic boundary
conditions.  Unlike some stencil problems, e.g., the Jacobi finite
difference solver for partial differential equations, in which the
data for the new time step is completely based on the previous time
step, the checkerboard decomposition allows the spin glass simulation
to proceed with two consecutive update phases. Only half of the spins are
updated in each of the phases. This unconventional stencil, associated
with the checkerboard decomposition, leads to a stride-2 memory
reference pattern and presents a more challenging memory optimization
problem compared to the stride-1 pattern of typical stencils problems.

\subsubsection{Allocation}

\begin{figure}[!h]
  \centering
\subfigure[Unified]{\includegraphics[width=0.15\textwidth]{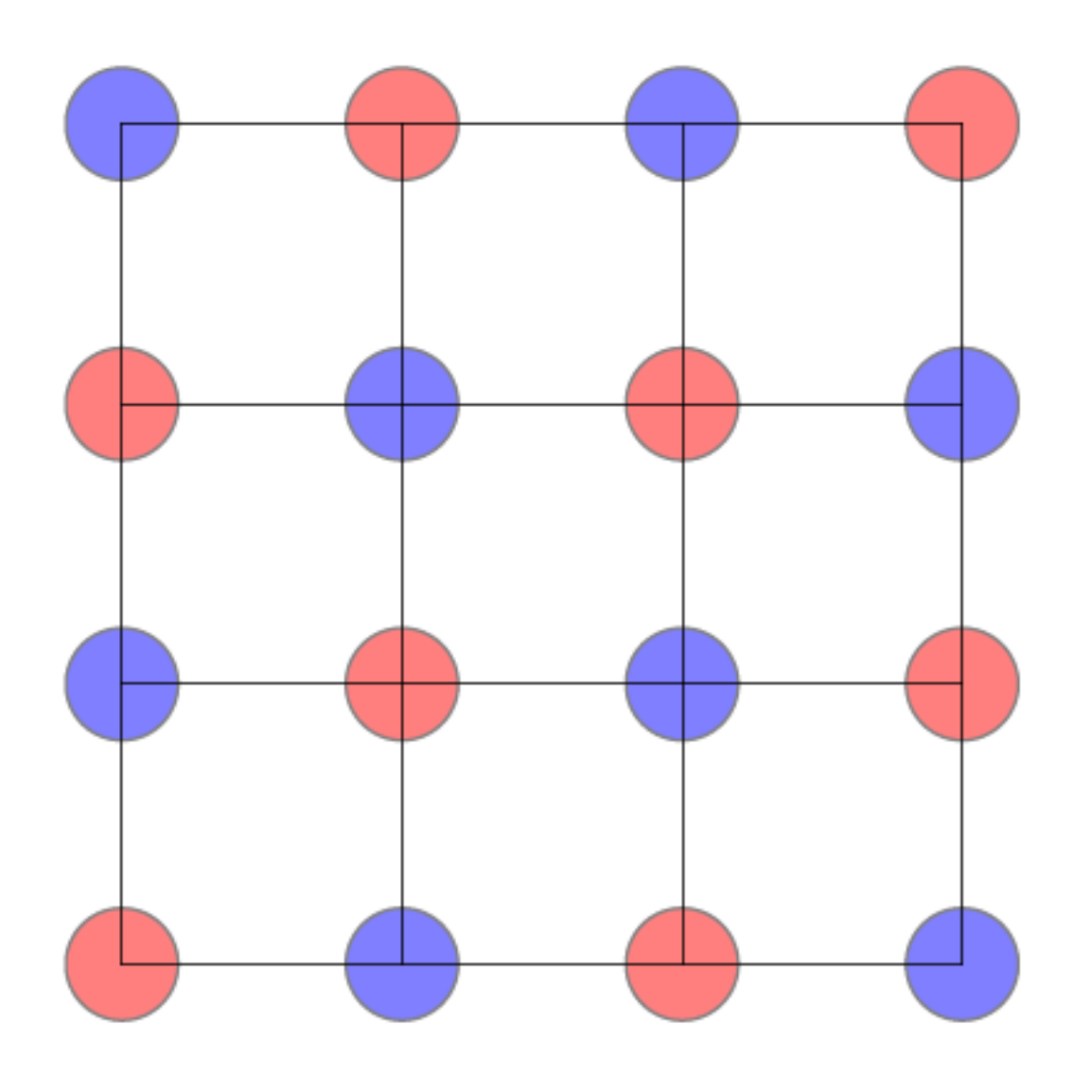}}\hspace{1.4cm}
\subfigure[Separated]{\includegraphics[width=0.173\textwidth]{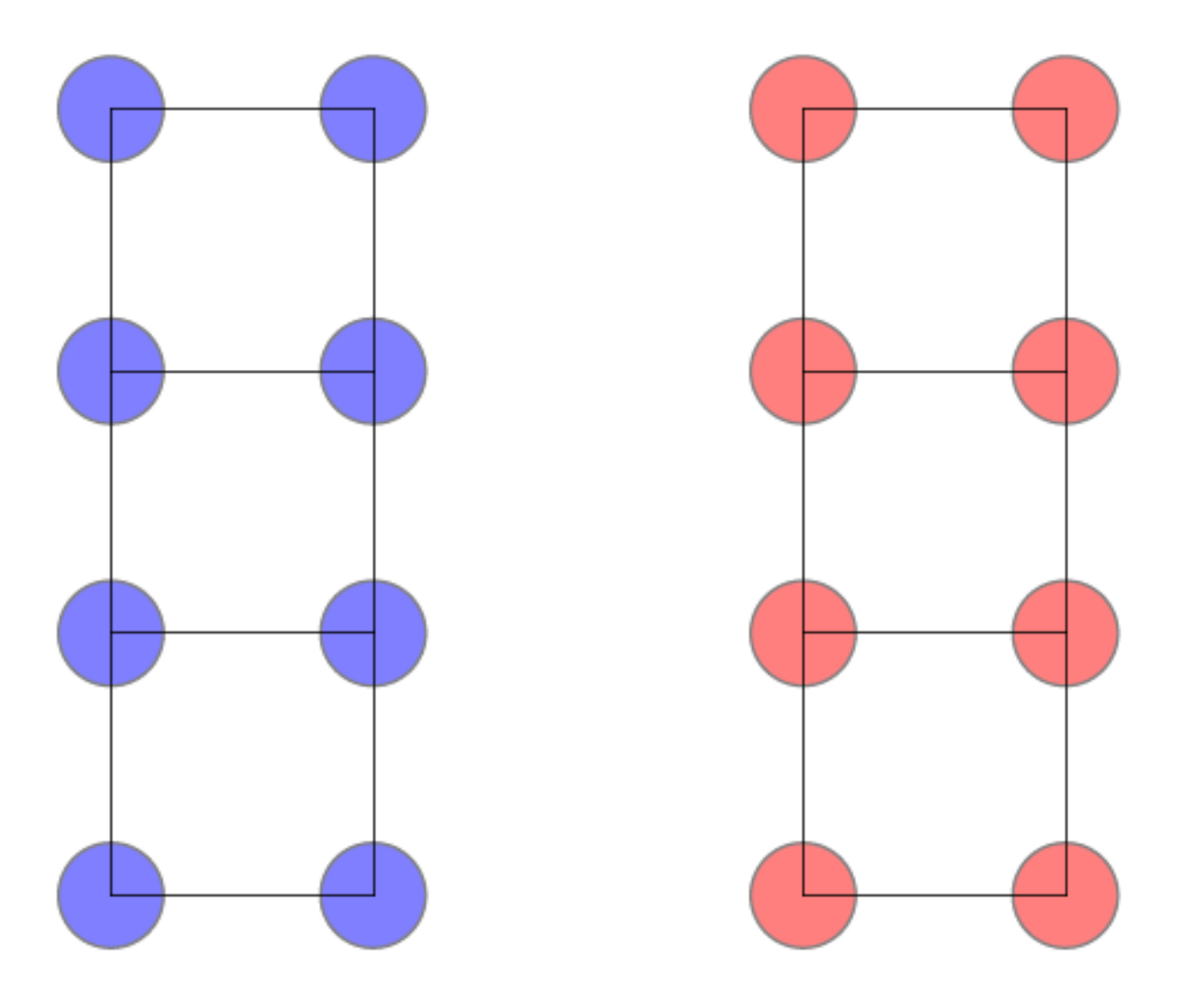}}
  \caption{Unified and separated memory allocation schemes.  The unified scheme stores the entire checkerboard
  lattice together. The separated scheme breaks the memory associated with each sublattice into
  separate continuous blocks of memory.}
\label{fig_alloc12}
\end{figure}
\begin{figure}[!h]
  \centering
  \includegraphics[width=0.4\textwidth] {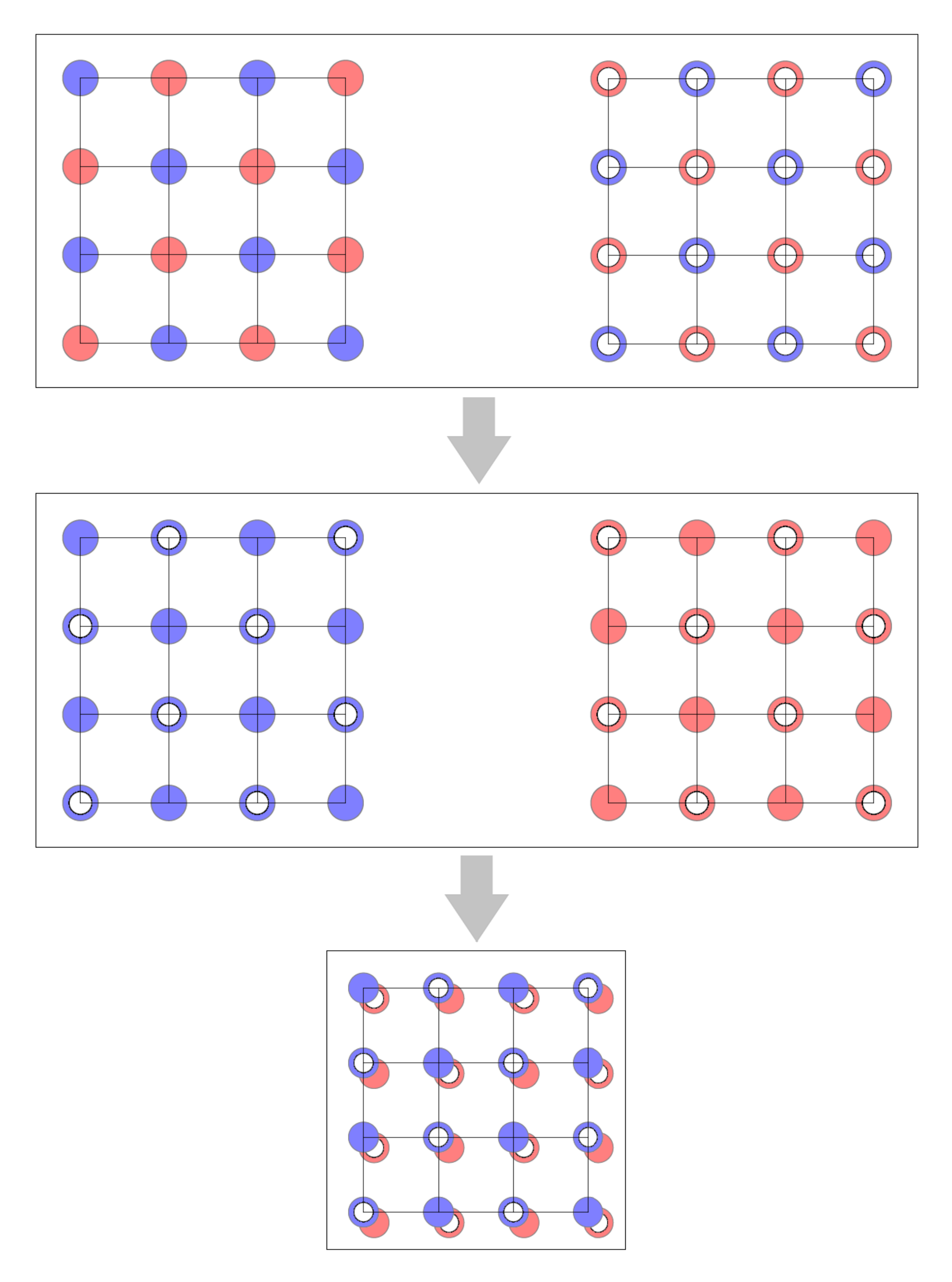}

  \caption{The shuffled allocation mixes and integrates two lattices,
shown on the top of the figure. The first transformation is taking
the A sublattice on the left (blue dots) and the B sublattice
on the right (blue circles) to construct an intermediate lattice
(middle left figure of blue color).
Another intermediate lattice of red color is constructed similarly.
We then integrate these two intermediate lattices together, which
occupy different bit positions under the compact multispin coding scheme (Section \ref{section_msc}).
By using one integer lattice instead of two, we avoid doubling the memory consumption.
Also, the memory access pattern is identical to that of the 7-point 3D Jacobi stencil.
}

\label{fig_alloc3}
\end{figure}

We propose three different schemes to address this problem.
\begin{enumerate}
\item The {\bf Unified} allocation (Figure \ref {fig_alloc12}(a))
  stores the checkerboard lattice in its native way as a single piece.
\item The {\bf Separated} allocation (Figure \ref{fig_alloc12}(b))
  breaks the sub-lattices into two chunks stored separately.
\item The {\bf Shuffled} allocation \cite{2008CoPhC.178..208B} (Figure
  \ref{fig_alloc3}) mixes and integrates two temperature replicas,
  so that the memory access pattern is now identical to the
  conventional stencil.  This is done by mixing the two temperature
  lattices in such a way that all the A sublattice spins from
  temperature 1 and the B sublattice spins from temperature 2 are
  packed together in the memory associated with one lattice. When the
  spins are being updated on this lattice, they are all independent
  of each other. They can be considered sequentially and
  continuously written into memory. Since there is no gap between each
  memory write, this should theoretically enhance the memory access
  speed.
\end{enumerate}

The performance comparison on Table \ref{table_allocation} suggests that the
separated allocation is inferior due to its significantly lower memory
performance. This is because of the more complicated control flows in
the code. 
Overall, the unified allocation provides the best memory performance
in terms of time spent for each spin and is used in our
implementation.

\begin{table}[h]
\begin{small}
  \begin{tabular*}{\hsize}{@{\extracolsep{\fill}}lccc@{}}
    \hline
                                 & Unified       & Separated     & Shuffled\\
    \hline
    Bandwidth(GB/s)              & 645.1         & 279.0         & 832.6\\
    Time per transaction (ps)    & 49.608        & 107.756       & 38.432\\
    Spins per transaction        & 24            & 24            & 16\\
    Time per spin (ps)           & 2.067         & 4.345         & 2.402\\
    \hline
  \end{tabular*}
\end{small}
  \caption{Performance comparison of the unified/separated/shuffled storage allocation schemes for
  a $16^3$ lattice. The definition of  a transaction is a sequence of  7 loads and a store that serve the spin update.}
\label{table_allocation}
  \end{table}

\subsubsection{Tiling for the Multispin Coding Lattice}
The basic idea of multispin coding (MSC) is to present many binaries or short vectors in a longer 
packed word. For example, Ising spins may be stored with a single bit per spin, with 0 being spin 
down and 1 being up.  In our particular implementation, we also encode the 4 bit string of one site's 
spin-flip probability table's row index (section \ref{section_prob}) into an integer word. 
MSC \cite{PhysRevLett.42.1390,Zorn-Hermann-Rebbi-1981} yields a more efficient way of calculating local energies 
($E$) and reduces the memory required for the spin configurations.  This packing prevents the Arithmetic 
Logic Unit, which performs integer arithmetic and logical operations, and the memory bandwidth from 
being under utilized. Also, a memory transaction (7 loads and 1 store) can serve the calculation of 
multiple spins, which helps improve the relative memory performance.

The usual practice for a single lattice MSC is integrating a line of
spins into an integer. We denote this conventional method as
Synchronous Multispin Coding ({\bf SMSC}). For the simulation of spin
glass models, the temperature replicas provide an alternative approach
with a different memory layout.  One can pack the spins at a
specific site but at different temperature replicas into an integer;
we call this the Asynchronous Multispin Coding ({\bf AMSC}). The main
idea of these two multispin coding schemes are:
\begin{itemize}
\item {SMSC}: A packed word stores the spins from a single replica, but
  different sites.
\item {AMSC}: A packed word stores the spins belonging to different
  temperature replicas of the same site.
\end{itemize}
We find the ASMC scheme to be more efficient. Its storage consumption
is small enough to fit in the GPU shared memory. Furthermore, AMSC's
index system is more straightforward, thereby simplifying
optimization.  The performance of these different MSC schemes is
described below.  Here, we briefly discuss how the words
associated with either scheme are organized into memory.

\begin{figure}[!h]
  \centering
  \subfigure[Global Memory]{\includegraphics[width=0.19\textwidth]{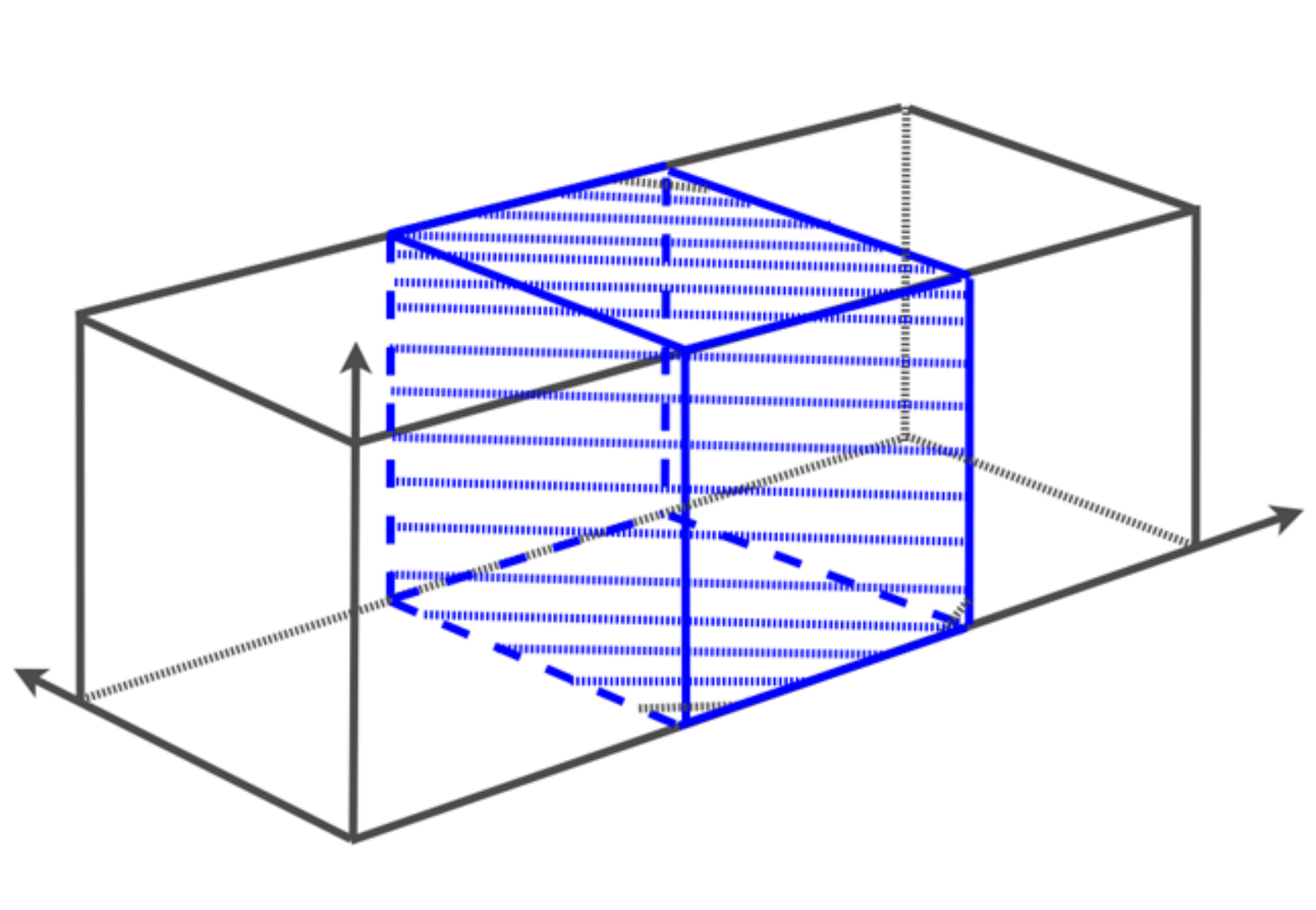}}\hspace{0.2cm}
  \subfigure[Shared Memory]{\includegraphics[width=0.12\textwidth]{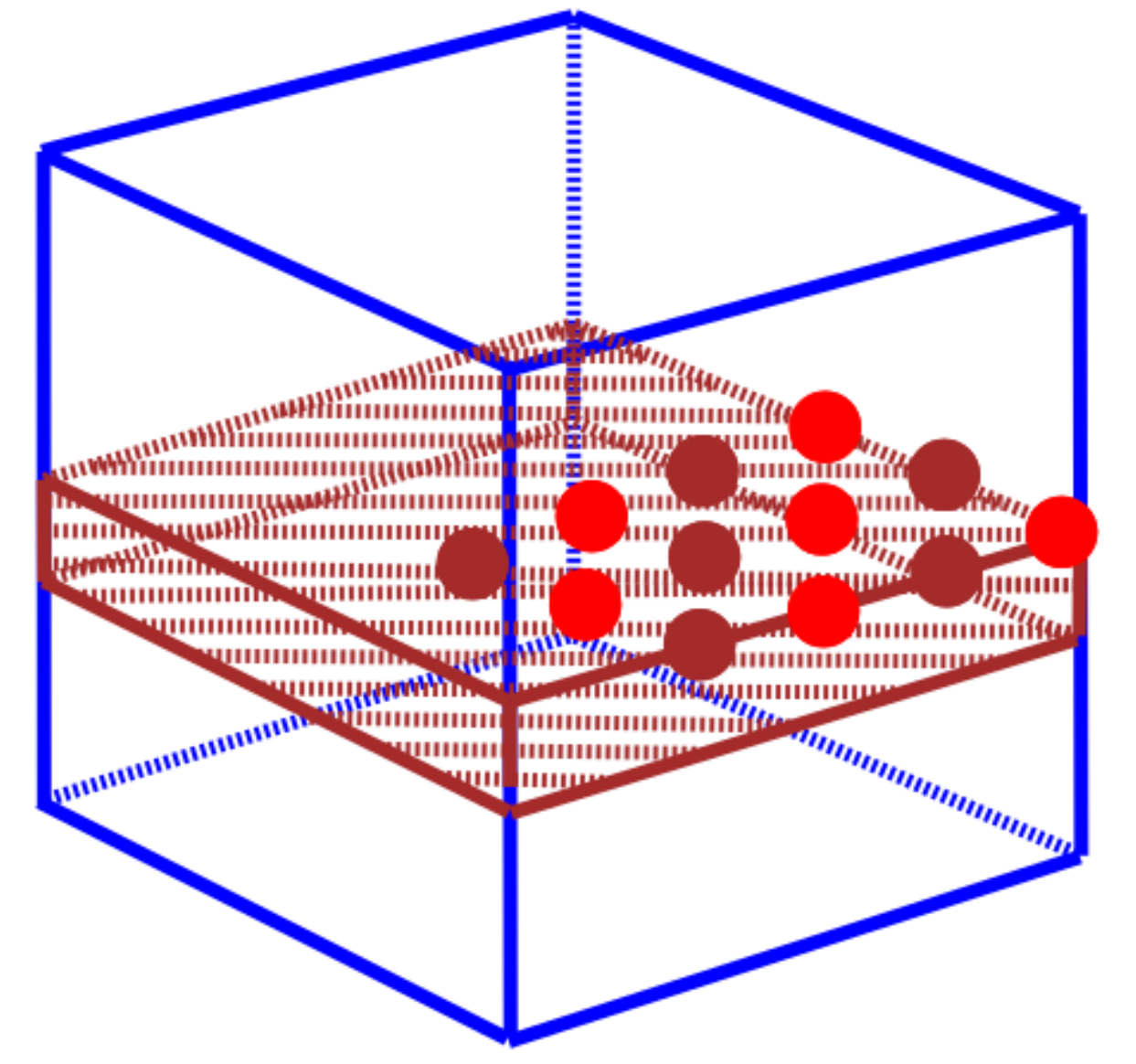}}\hspace{0.2cm}
  \subfigure[Registers]{\includegraphics[width=0.1\textwidth]{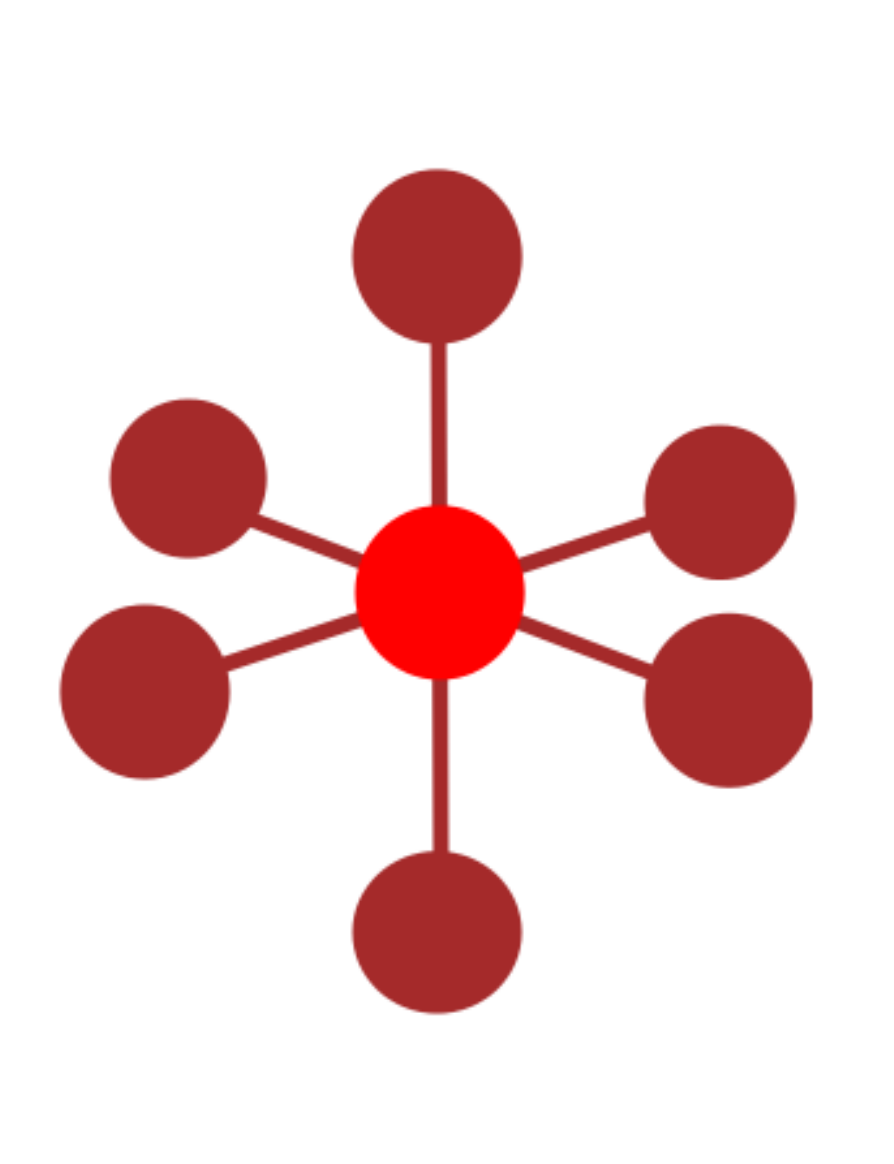}}

  \caption{Memory tiling.  The global memory may hold several integer cubes (including one integer
per lattice site) if there are more than 32 temperature replicas.  The shared memory scratchpad holds 
the working set of an entire integer cube (no larger than $4 \times 16^3 = 16KB$).  The registers hold the
data needed for local spin updates.}
  \label{fig_tile}
\end{figure}

Three levels (Figure \ref{fig_tile}) of the memory hierarchy are employed
that reflect the GPU memory architecture of global memory, shared
memory and registers:
\begin{itemize}
\item {\bf Level 1:} The main data resides in the GPU global
  memory. Due to the limitation that a 32 bit integer represents at
  most 32 spins, we may need multiple integer cubes (with an integer
  cube including one integer per site on the cubic lattice) if there
  are more than 32 temperature replicas.

\item {\bf Level 2:} The shared memory scratchpad holds the working
  set of an entire integer cube (no larger than $4 \times 16^3 =
  16KB$).  The data transfer between global and shared memory is quite
  modest because we do not need to switch to another integer cube
  until the Monte Carlo and parallel tempering swaps within the
  temperature replicas contained within the current cube are
  exhausted.

\item {\bf Level 3:} The GPU threads scan the shared memory scratchpad
  for two consecutive sublattices and load the data into
  registers. The threads are organized as multiple layers of 2D
  plates.  We observe the optimal thread configurations are two or
  four layers ($16^2/2 \times 2 = 256$ or $16^2/2 \times 4 = 512$).
\end{itemize}

\subsection{Optimizing the Computation}

\label{section_compute}

We may take advantage of the MSC mapping of the spins onto bits to
dramatically reduce the number of floating point operations needed by
the Monte Carlo parallel tempering calculations.  
For example, we may use a bitwise XOR ($\oplus$) as
opposed to multiplication to calculate the energy. In the equations
below, we denote the variables in the original notation with a
superscript $^o$, and variables without superscripts are used in the
transformed notation. The variables $S$,
$J$, $e$ and $E$ stand for
spin, spin coupling, bound energy and local energy respectively.


\[
S^o \in \{-1, 1\}, J^o \in \{-1, 1\}
\]
\[
E^o_{i} = \sum_{j} S^o_i \times J^o_{ij} \times S^o_j, E^o_{i} \in \{-6, -4, -2, 0, 2, 4, 6\}.
\]
\[
S \in \{0, 1\}, J \in \{0, 1\}
\]
\[
E_{i} = \sum_{j} S_i \oplus J_{ij} \oplus S_j, E_{i} \in \{0, 1, 2, 3, 4, 5, 6\}.
\]
Note that local energy $E^o_{i} $, the energy of a spin $i$ in the field of its nearest neighbors, can only take 
one of seven values as indicated.  

The computation is composed of four steps:
\begin{enumerate}
\item Energy:
Compute the bound energy ($e$) and the spin's local energy ($E$).
\begin{align}
\label{eq:e}
\begin{split}
e_{ij} = S_i \oplus J_{ij} \oplus S_j\\
E_{i} = \sum_{j} e_{ij}
\end{split}
\end{align}

\item Probability:
Compute the flip probability ($P$) for the Metropolis Monte Carlo,
where the temperature ($T$) is an input parameters.
\begin{align}
\begin{split}
\label{eq:p}
E^o = 2 \times E - 6 \\
S^o = 2 \times S - 1 \\
P = \mr{exp} (2 \times (\frac{1}{T} \times E^o + h \times S^o))
\end{split}
\end{align}

\item Rand:
Generate a random number ($R$).

\item Compare:
Compare and update spins.
\begin{equation}
\label{eq:r}
S = (P < R) \oplus S.
\end{equation}
\end{enumerate}

\subsubsection{Probability Look-up Table}

\label{section_prob}

\begin{figure}[!h]
  \centering
  \includegraphics[width=0.45\textwidth]{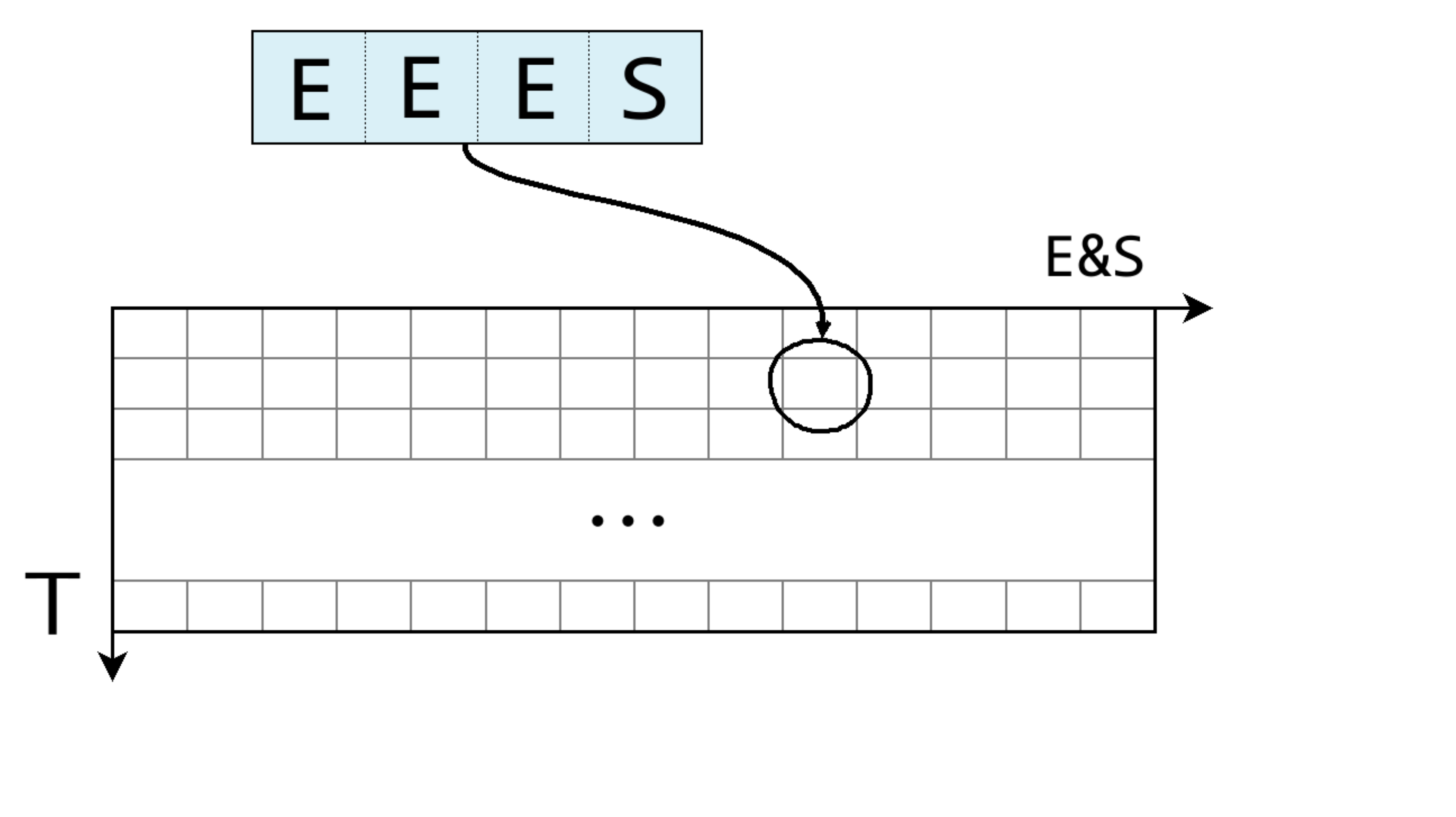}
  \caption{The organization of the probability look-up table.}
\label{fig_table}
\end{figure}

Eq. \ref{eq:p} expresses the straightforward yet expensive method to generate the spin flip 
probabilities. However, since the number of input/output values is finite (i.e., combinations of 7 
possible local energies $E$, 2 spins $S$, and no more than 32 temperatures $T$), a better solution 
is to deploy a pre-calculated look-up table. The table is a two-dimensional matrix (Figure \ref{fig_table}), 
with $T$ as the row index and $(E \times 2 + S)$ as the column index.  The column index, as the 
combination of $E$ and $S$, requires 4 bits for the address space. The maximum storage consumption 
of the table is 16 KB, assume that we have 32 rows times 14 columns times 4 bytes per entry (again, 
assume 32 temperature replicas).  When a parallel tempering swap between two replicas at temperatures 
$T_{i}$ and $T_{j}$ is accepted, the two corresponding rows in the table are swapped.

\begin{figure}[!h]
  \centering
  \includegraphics[width=0.45\textwidth]{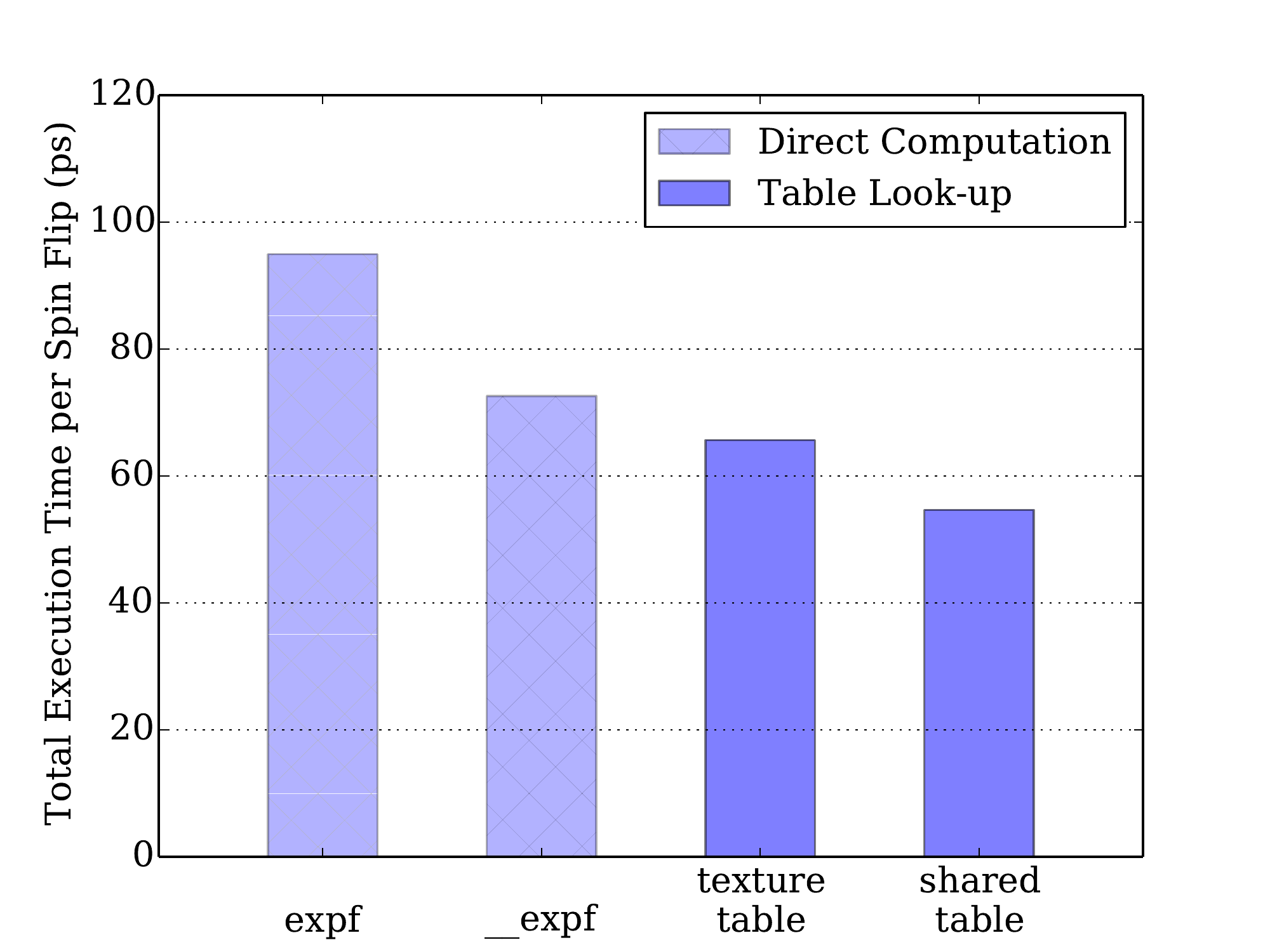}
  \caption{A comparison of the overall time consumed per spin flip 
using four different methods to compute the exponential probability in 
Eq.~\ref{eq:p} as described in the main text. 
The experiment is done for a $16^3$ lattice, fp32 CURAND and AMSC1. 
  No parallel-tempering is performed.}
  \label{fig_perf_prob}
\end{figure}

We evaluate four different ways to calculate the probability in Eq.~\ref{eq:p} (Figure \ref{fig_perf_prob}): 
(a) using the floating point exponential function from the math library, 
(b) using a less accurate GPU specialized exponential intrinsic function, 
(c) using the texture memory to store a table, and 
(d) a shared memory table. 
The result shows that an optimal table look-up saves close to half of the total computation time
compared to direct computation of the probabilities. In addition, 
the shared memory table outperforms the texture memory table. This is
because GPU threads are simultaneously computing on the same
temperature replica, and are therefore accessing the same row of the
table. This avoids bank conflicts, so that the high bandwidth and low
latency performance potential of the shared memory is fully exploited. 

\subsubsection{Random Number Generator}

\label{section_rng}

The simulation requires uniformly distributed random numbers between
zero and one. However, due to the fact that pseudo random number
generators (RNGs) manipulate integer values internally, directly using integer
return values from the RNG provides higher performance and preserves identical
precision.  As a consequence, we convert the pre-generated
probabilities from single precision floating point numbers to 32 bit
unsigned integers.

\begin{figure}[!ht]
  \centering
  \includegraphics[width=0.45\textwidth]{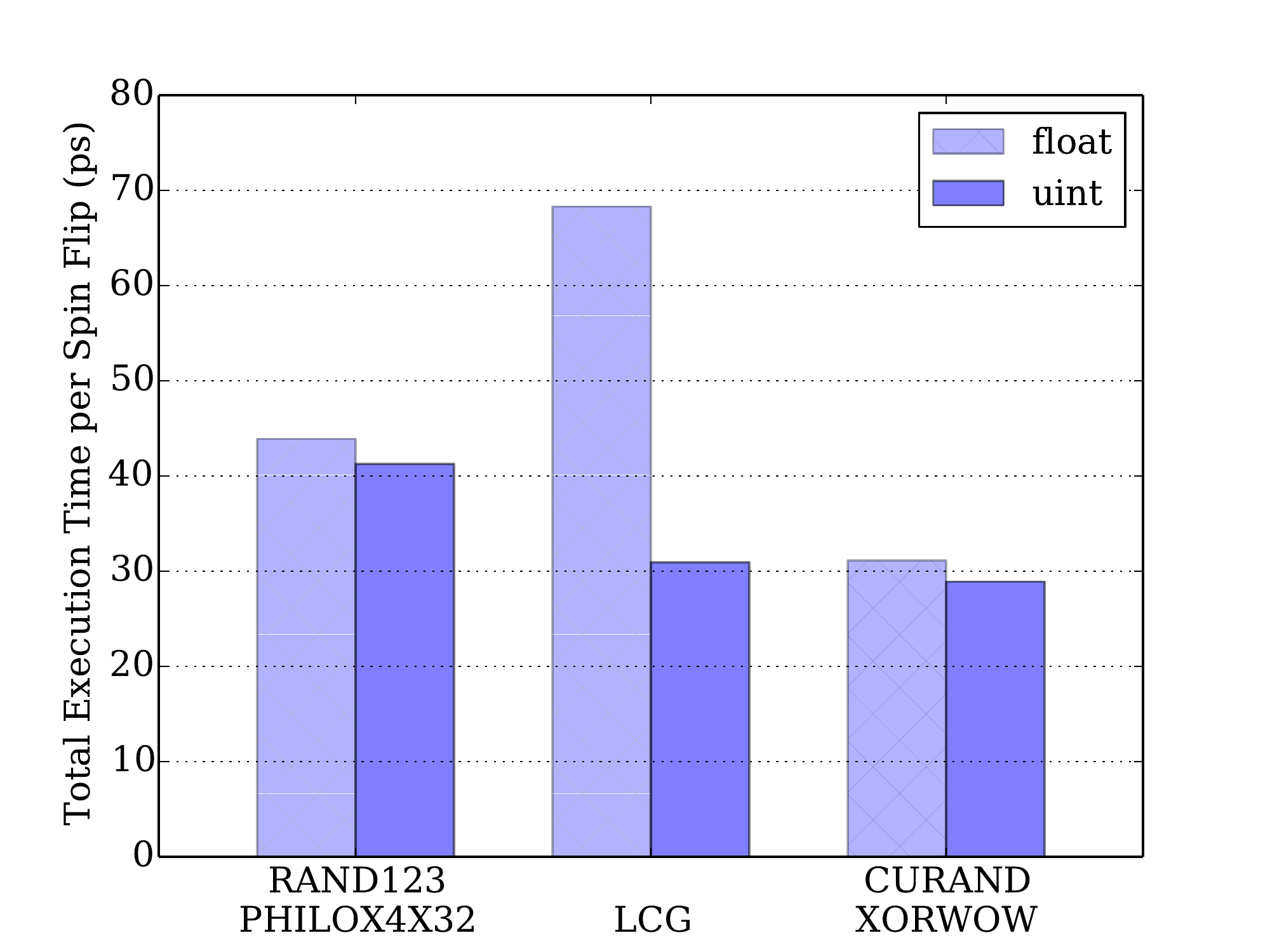}
  \caption{A comparison of the overall time required per spin flip using different random number generators. 
  The experiment used a $16^3$ lattice, a shared memory probability table and CAMSC. No 
  parallel-tempering is performed. The loop that consumes random numbers has been unrolled four 
  times to match the four return values of rand123 philox4x32\_7.} 
  \label{fig_rng}
\end{figure}

We evaluated three random number generators: (i) NVIDIA CURAND library
of XORWOW algorithm \cite{curand}, (ii) rand123
\cite{Salmon:2011:PRN:2063384.2063405} philox4x32\_7 (version 1.06),
and (iii) our implementation of a multi-threaded 32 bit linear
congruential generator (LCG).  We decide to adopt CURAND due to its
higher performance (Figure \ref{fig_rng}) and quality
\cite{2012arXiv1204.6193M}.


\subsubsection{Multispin Coding}

\label{section_msc}

\begin{figure*}[ht]
  \centering
  \subfigure[AMSC1]{\includegraphics[width=0.267\textwidth]{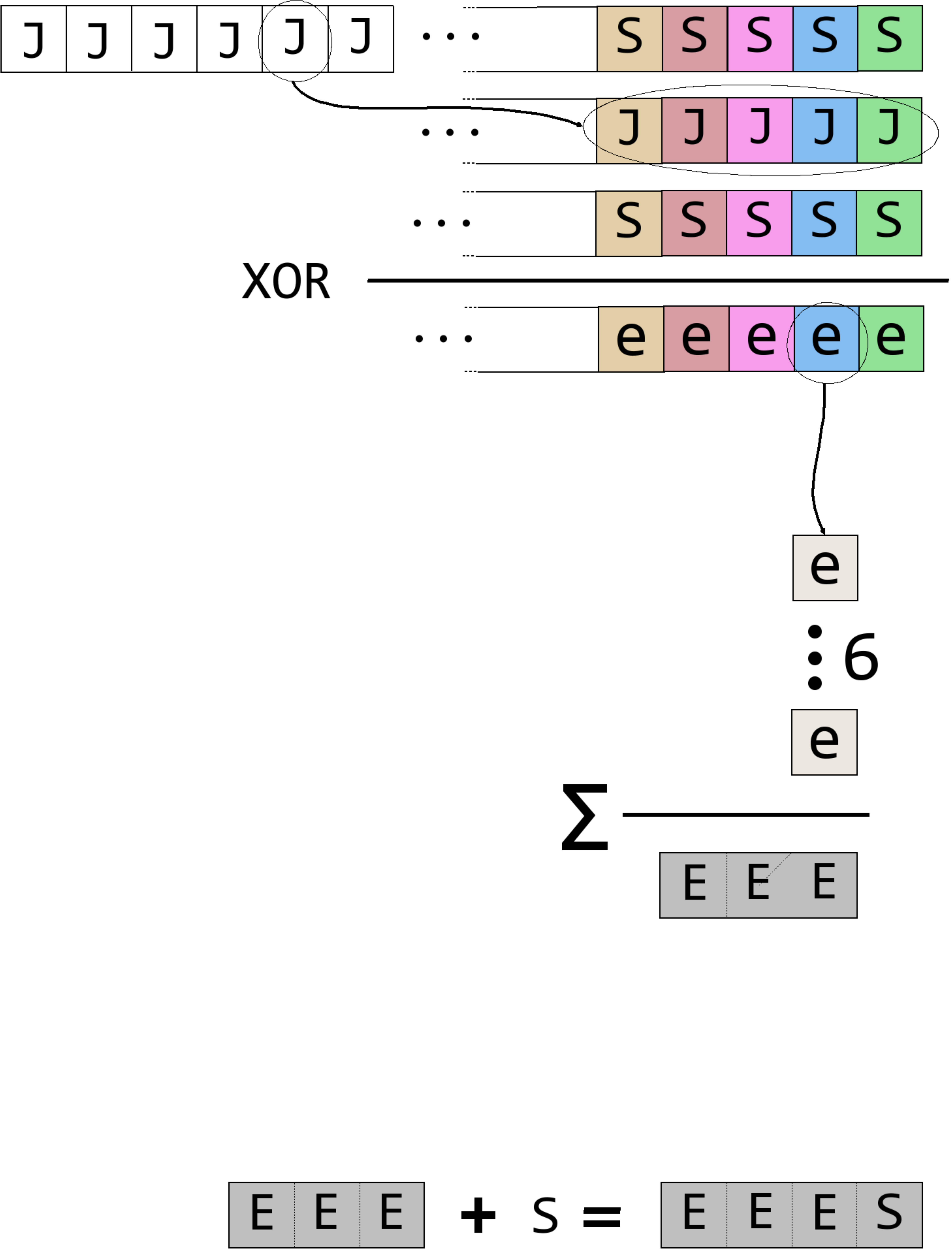}}\hspace{2.0cm}
  \subfigure[AMSC3]{\includegraphics[width=0.31\textwidth]{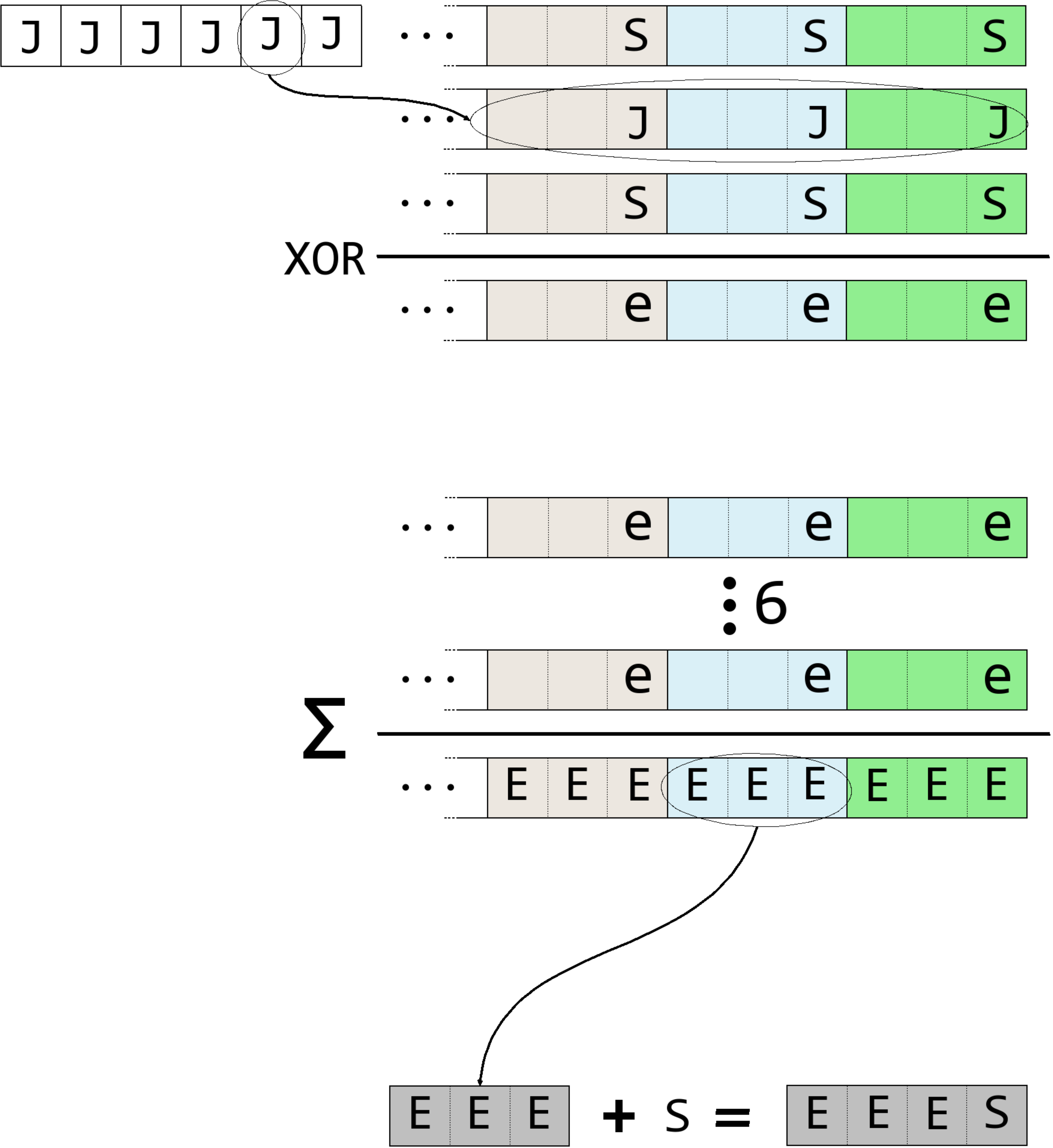}}\\
\caption{This figure demonstrates the computation of $(E \times 2 +
  S)$ for the purpose of accessing the probability look up table with
  the deployment of two variations of Asynchronous Multispin Coding
  (AMSC).  Each line in the figure represents an integer, each box of
  a line represents a bit, and boxes of the same color represent a
  segment that hold a variable from one of the temperature replicas.
  We give the name AMSC1 and AMSC3 for these two AMSC schemes
  according to their segment width.  Unlike the AMSC1, the AMSC3
  scheme reserves three bits for each segment, and is a less dense
  storage format.  For the calculation of the local energy, we need
  two spins ($S$) and the coupling ($J$) between them. The $J$ bits
  and $S$ bits are integrated in the same integer, so that we can
  fetch both the coupling and the spins using only one memory
  transaction. The local energy ($e$) of each bond 
  can be calculated by performing two XOR operations. The
  total local change of energy ($EEE$) is the sum of the contributions
  from all six nearest neighbors. Since $EEE$ requires three bits for
  storage, the AMSC1 scheme compute each segment sequentially to
  avoid overflow, while the AMSC3 scheme can compute multiple segments in
  parallel.  After we obtain $EEE$ in three-bit format, we combine it
  with the spin state ($S$) by doing string concatenation.}
\label{fig_msc1}
\end{figure*}

\begin{figure*}
\centering
 \subfigure[AMSC4]{\includegraphics[width=0.284\textwidth]{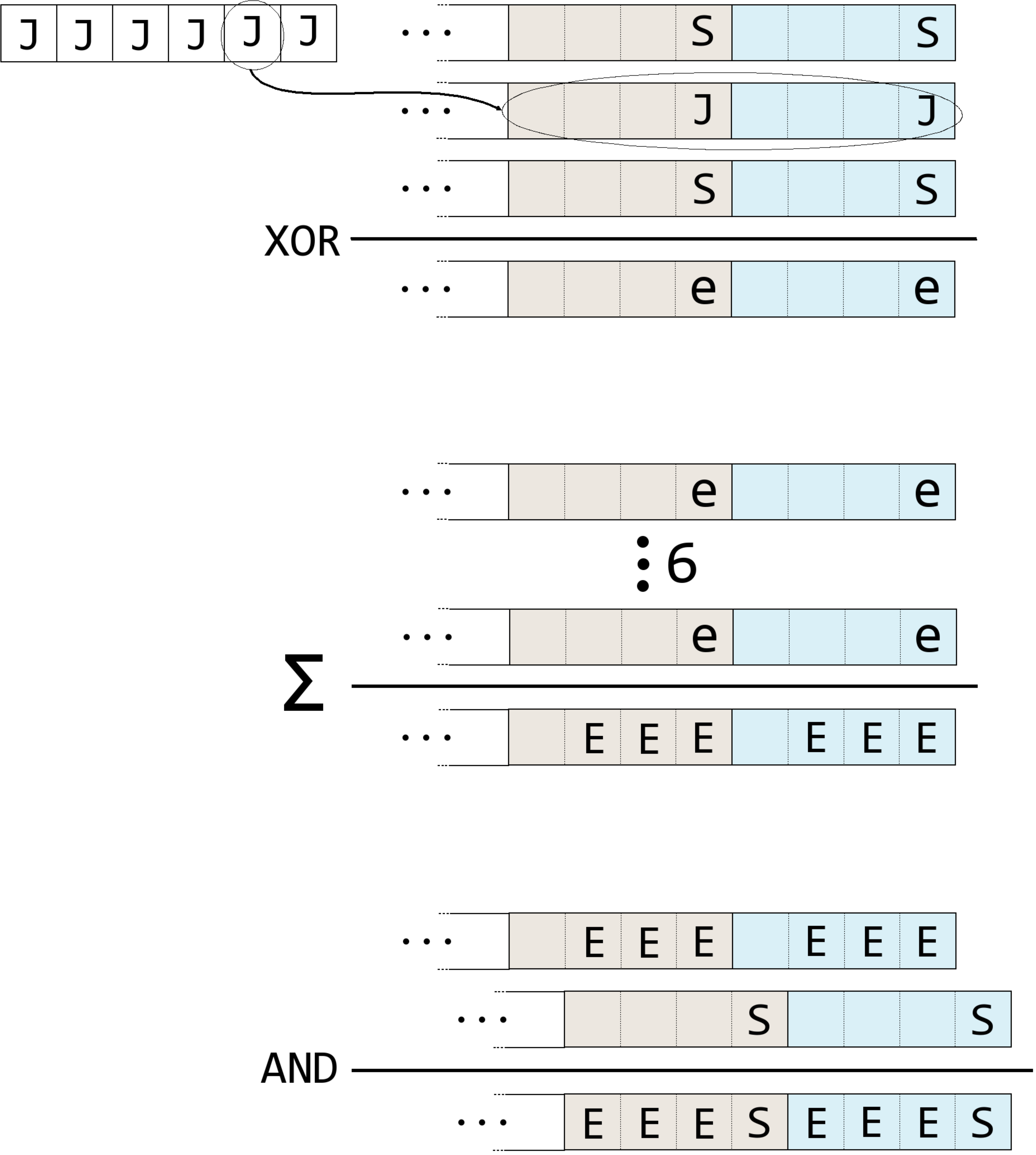}}\hspace{2.0cm}
 \subfigure[CAMSC]{\label{fig_camsc}\includegraphics[width=0.28\textwidth]{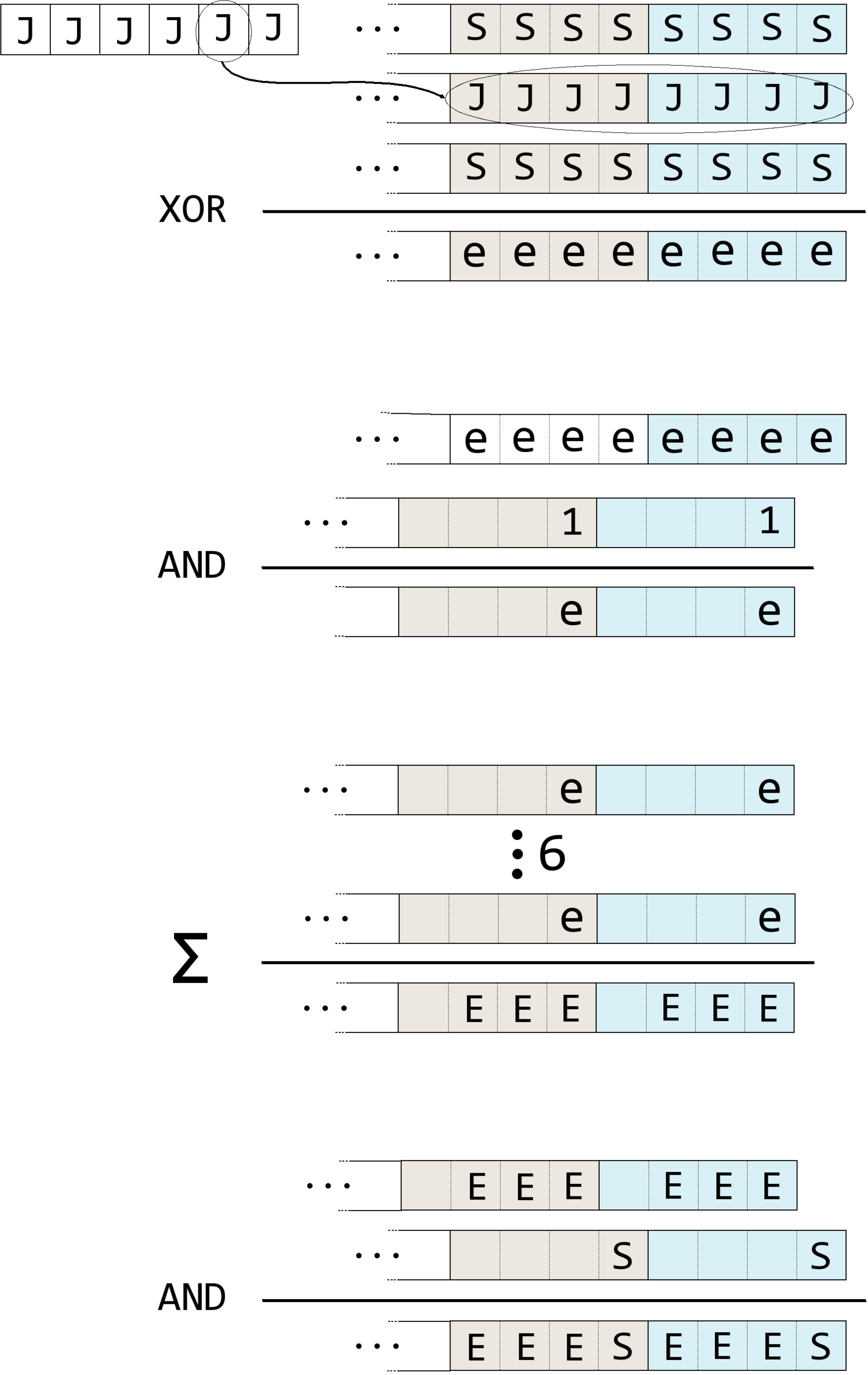}}
\caption{This figure demonstrates how the AMSC4 and CAMSC schemes help
  exploit bit-level parallelism in computing $(E \times 2 + S)$.
  Similar to that of the AMSC1 and AMSC3 (see the text and the caption
  of Fig.~\ref{fig_msc1}), the XOR operations and summation over six
  nearest neighbors produces the total local energy ($EEE$).  However,
  since we reserve four bits for each segment, and is capable of
  holding one more bit over $EEE$, the string concatenation of $EEE$
  with $S$ can now be vectorized.  The difference between CAMSC and
  AMSC4 is that $S$ and $J$ are stored in a more compact format. With
  such a design, CAMSC avoids waste of space and provides much better
  parallelism in computing $e$. }
  \label{fig_msc2}
\end{figure*}

We have briefly described the Multispin Coding (MSC). We have developed the 
Asynchronous Multispin coding (AMSC) as a more efficient
alternative to the conventional Synchronous Multispin Coding (SMSC)
for calculating the local energies ($E$), generating the 4 bit string
for the column index of the spin-flip probability table (section
\ref{section_prob}), and optimizing the memory bandwidth
utilization. In our particular GPU implementation, we use four byte
unsigned integers, which hold up to 32 bits, as a packed word. Each
spin, denoted as 0 or 1, takes only one bit of this packed word. Thus, the
calculations in Eq. \ref{eq:e} can be vectorized via bit-wise
operations. We integrate the $J$ bits with the $S$ bits in the same
integer, so that we can fetch both the coupling and the spins in only
one memory transaction. We then multiply the coupling with a bit-mask
to match the pattern of $S$, and calculate the bond energy with
bit-wise XOR operation. The next step is to add the six bond energies
around a spin to obtain the local energy.  To vectorize this process
we need to reserve empty bits to avoid overflow, since the local
energy takes 3 bits of storage. In this way, each spin, together with
the empty bits reserved for calculation, constitute a virtual segment.
We derived three variations of AMSC with different segment width of 1, 3 and 4, denoted as 
AMSC1, AMSC3 and AMSC4 respectively. In AMSC1 and AMSC3, some calculations are 
sequentialized to avoid overflow. Figures \ref{fig_msc1} and \ref{fig_msc2} demonstrate how the  
the different variations of MSC parallelize the computations in Equations~\ref{eq:e}, \ref{eq:p} 
and~\ref{eq:r}. 

\begin{figure}[!h]
  \centering
  \includegraphics[width=0.45\textwidth]{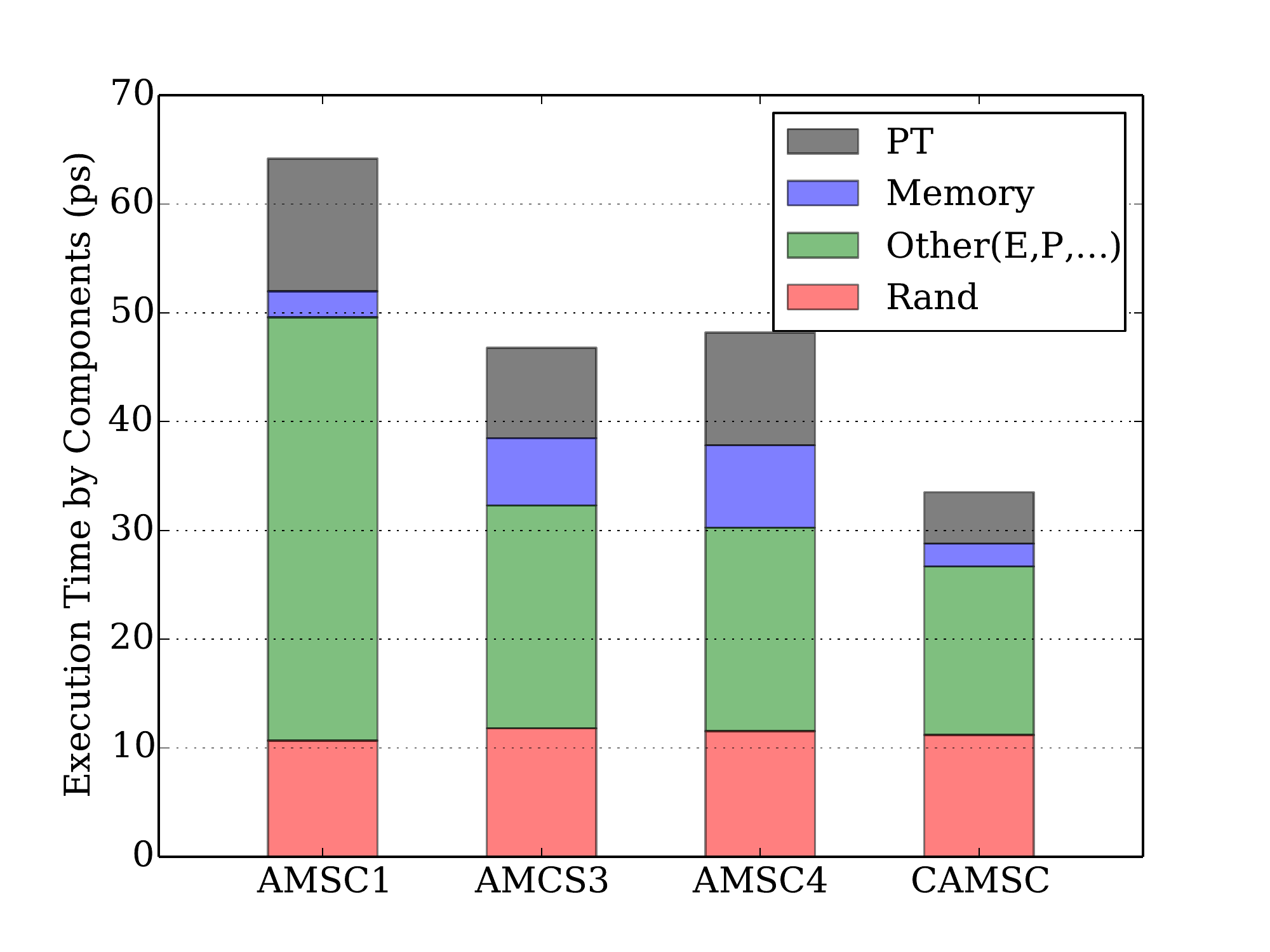}
  \caption{Comparing the performance using different multispin coding schemes. The experiment 
is done for a $16^3$ lattice, a shared memory probability table with integers and CURAND.
A parallel-tempering move is performed every 10 Metropolis single spin sweeps.}
  \label{fig_msc_perf}
\end{figure}

Figure \ref{fig_msc_perf} illustrates that AMSC3 and AMSC4 are favored over AMSC1 due to improved 
overall performance.  However, we also observe proportionally longer times for
the memory transactions. 
This demonstrates the limitation of the AMSC scheme: there does not exist an optimal segment width
that simultaneously provides the highest memory density, and the richest vectorization opportunities in computation.


To overcome the intrinsic limitation of AMSC, we propose a new scheme named Compact Asynchronous Multispin 
Coding (CAMSC).  We dynamically change the segment width to match the data range. 
Longer width is adopted for larger data to qualify the vectorization of computing multiple segments. 
For small range data, we use shorter width to avoid blank bits reservations. 
For example, we allocate 1 bit per segment for $S$ and $e$, and then expand to 4 bits when calculating $E$. 
The segment width expansion is implemented with shift and mask operations. 
Figure \ref{fig_camsc} demonstrates the procedures of CAMSC and how it differs from traditional AMSCs.
Our experiment (Figure \ref{fig_msc_perf}) shows 28.4\% performance improvement 
when we switch from AMSC3 (46.8 ps/spin) to CAMSC (33.5 ps/spin).


\section{Experimental Results}
\label{section_exp}

\subsection{The Platform Settings}
\label{section_platform}

Our development and performance evaluations are carried out on a
workstation with an Intel Core i7 x990 CPU and an NVIDIA GeForce GTX
580 GPU card. The GeForce GTX 580 is equipped with a Fermi architecture
GPU of 512 stream processors. We use Linux 2.6.32 x86-64, CUDA toolkit version 4.1
and gcc 4.4.6, and optimization flag -O2. We always configure the
GPU on-chip memory as 48KB shared memory plus 16KB L1 cache.

\subsection{Performance Evaluation}
\label{section_compare}
To evaluate the performance we use the time spent (in picoseconds) per spin flip proposal,
abbreviated as ps/spin (See eq. \ref{eq:tsf}).

When we study the equilibrium properties of a spin glass, the system
sizes that can be equilibrated within a reasonable time are not very
large. Therefore, we used $L=16$, or $N_\mathrm{spins}=4096$ as the
maximum system size. Meanwhile, to achieve efficient parallel
tempering moves, we set the number of temperature replicas to $N_T=24$ or $56$, 
and perform frequent parallel tempering moves (one parallel tempering move after
every $5$ to $10$ Monte Carlo sweeps). The typical number of Monte
Carlo steps required to equilibrate such a system is approximately
$10^7$.  Due to the huge sample-to-sample variation, a large number of
disorder realizations ($10^4$ or more) are usually required. However, since there is no
correlation among different realizations, we can scatter the jobs to
different GPU cards or nodes on a cluster.  On each of the cards we
only need $16$ to $64$ realizations to fully utilize all the
multiprocessors.

For benchmarking, we simulate 64 disorder realizations of the
EA model on a $16^3$ lattice with 24 temperature
replicas, and propose to swap adjacent temperatures every 10 Monte
Carlo sweeps.  We are able to complete $10^7$ Monte Carlo sweeps in 40 minutes. 
This wall time consists of the single spin flip Monte Carlo time,
the parallel tempering swap time, and the measurement time.
Discarding the measurements, the average computational speed is 33.5
ps/spin, for a single GPU device.
If we simulate without parallel tempering and serve all
temperature replicas with the same random number, we could obtain 17.6
ps/spin.  Generating random numbers consumes about one third of the
total simulations time, as shown in Figure~\ref{fig_msc_perf}.  We
believe we are approaching the limit of performance optimization. For
reference, our single thread CPU code (using AMSC4 without parallel tempering on a $16^3$ cubic lattice) runs at the speed of 14737
ps/spin; this represents a speed up of almost 440 for the GPU code over
the CPU code.

\begin{figure}[!ht]
  \centering
  \includegraphics[width=0.5\textwidth]{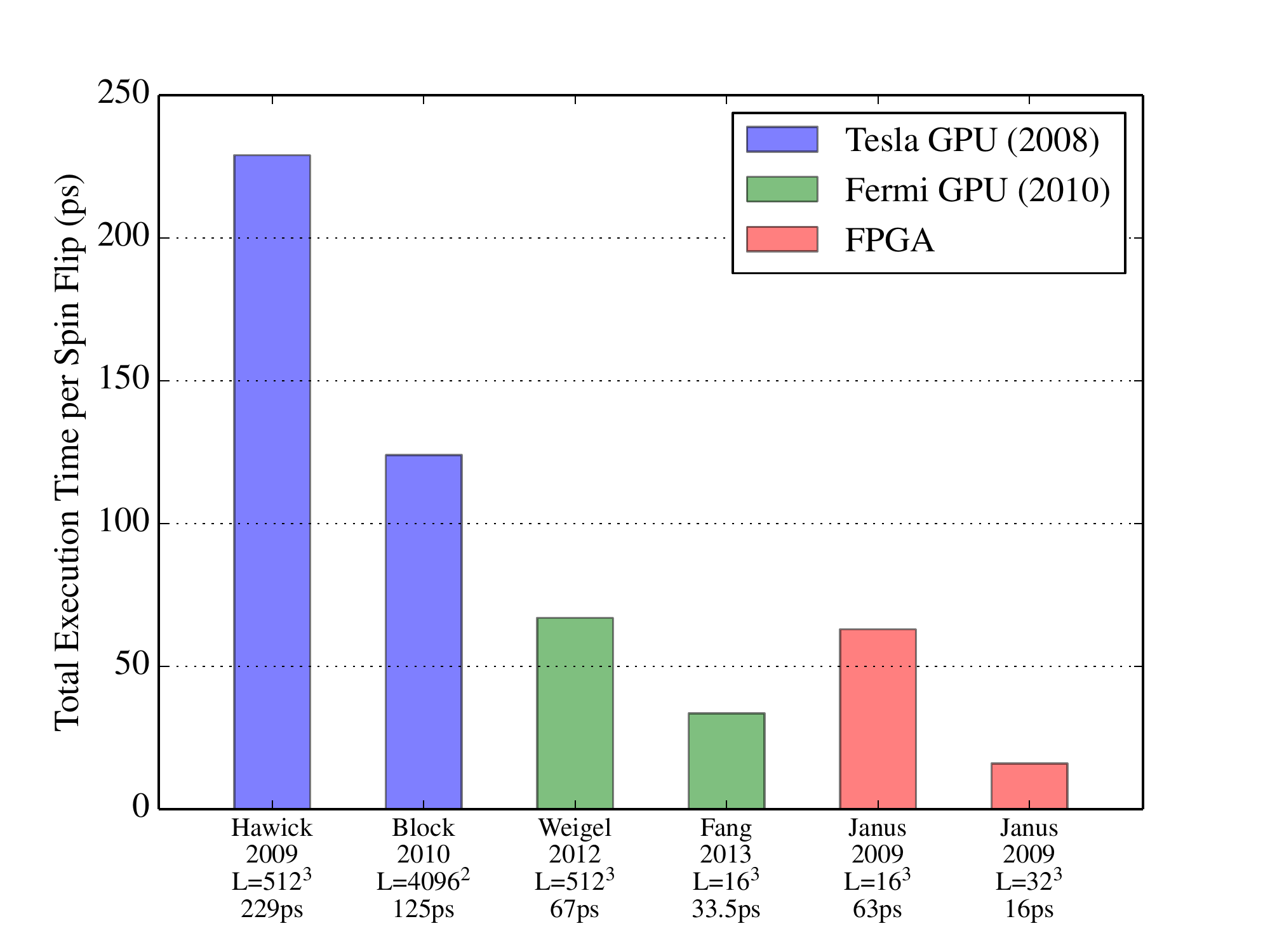}
  \caption{Performance comparison with other heterogeneous Ising model simulation programs.
\citet{CSTN-093} reports 4360.1 million Monte Carlo hits per second,
which equals to 229 ps/spin.
\citet{2010CoPhC.181.1549B} reports 7977.4 spin flips per microsecond,
which equals to 125 ps/spin.
}
  \label{fig_perf}
\end{figure}

Figure \ref{fig_perf} compares our implementation with similar
existing codes, where not all reference programs target at the
random frustrated Ising systems, present the external magnetic field, and feature parallel tempering.
Our program is substantially faster than any other GPU implementation
\cite{CSTN-093,2010CoPhC.181.1549B,doi:10.1142/S0129183112400025,Weigel:2012:PPS:2151219.2151631} for
small to intermediate system sizes. We are comparable to the
performance achieved by special-purpose FPGA
implementations\cite{2012arXiv1204.4134J}.

\subsection{Simulation Results}
\label{section_result}

\begin{figure}[ht!]
    \centering
    \includegraphics[width=0.5\textwidth]{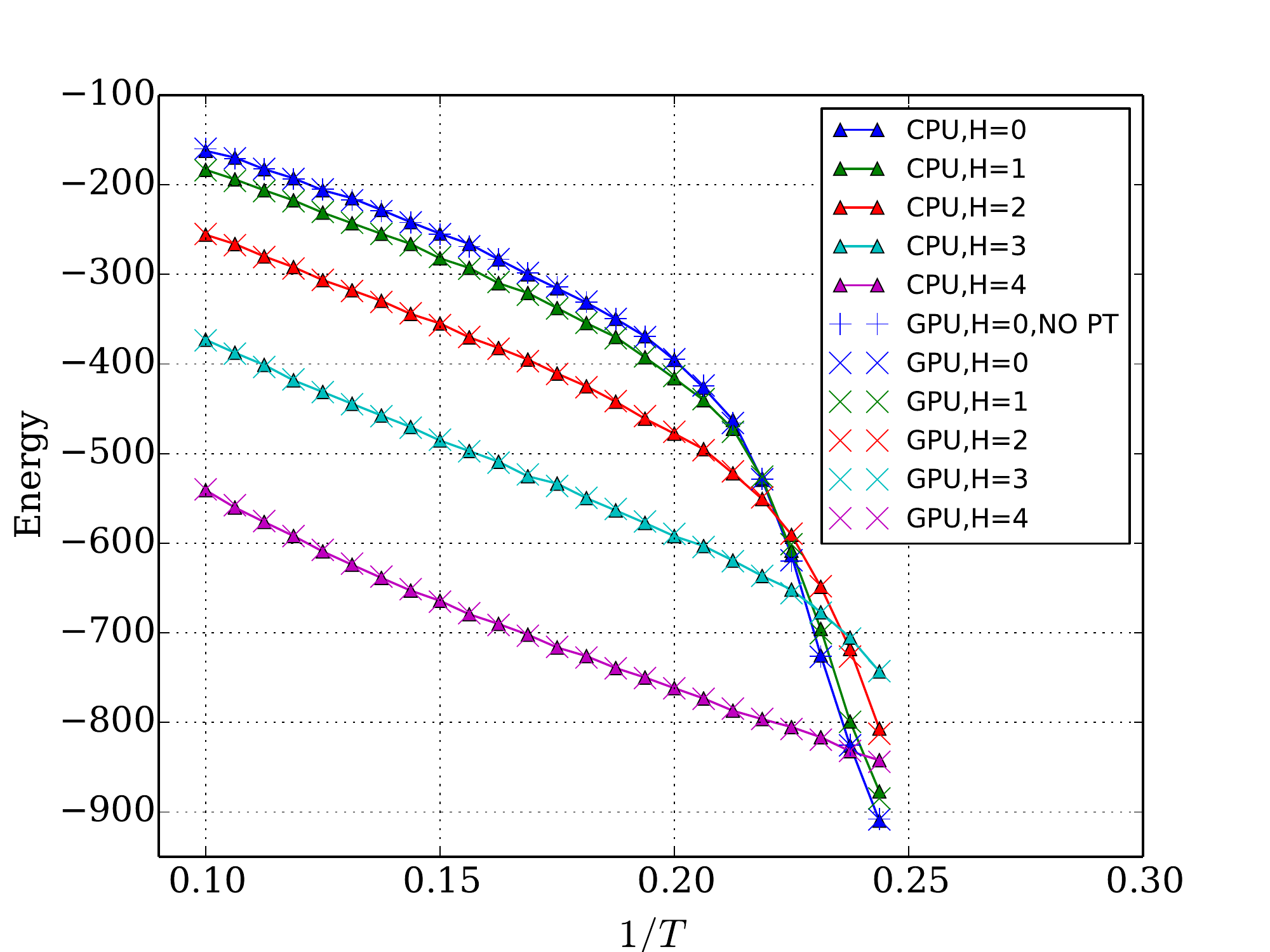}
    \caption{Comparing the total energy of the $16^3$ sites Ising model with nearest neighbors coupling $J=-1$, to CPU generated results. At each value of 
      the external field, the GPU results are nearly identical to the CPU results. }
    \label{fig:energy-ising}
    \end{figure}

\begin{figure}[ht!]
    \centering
    \includegraphics[width=0.5\textwidth]{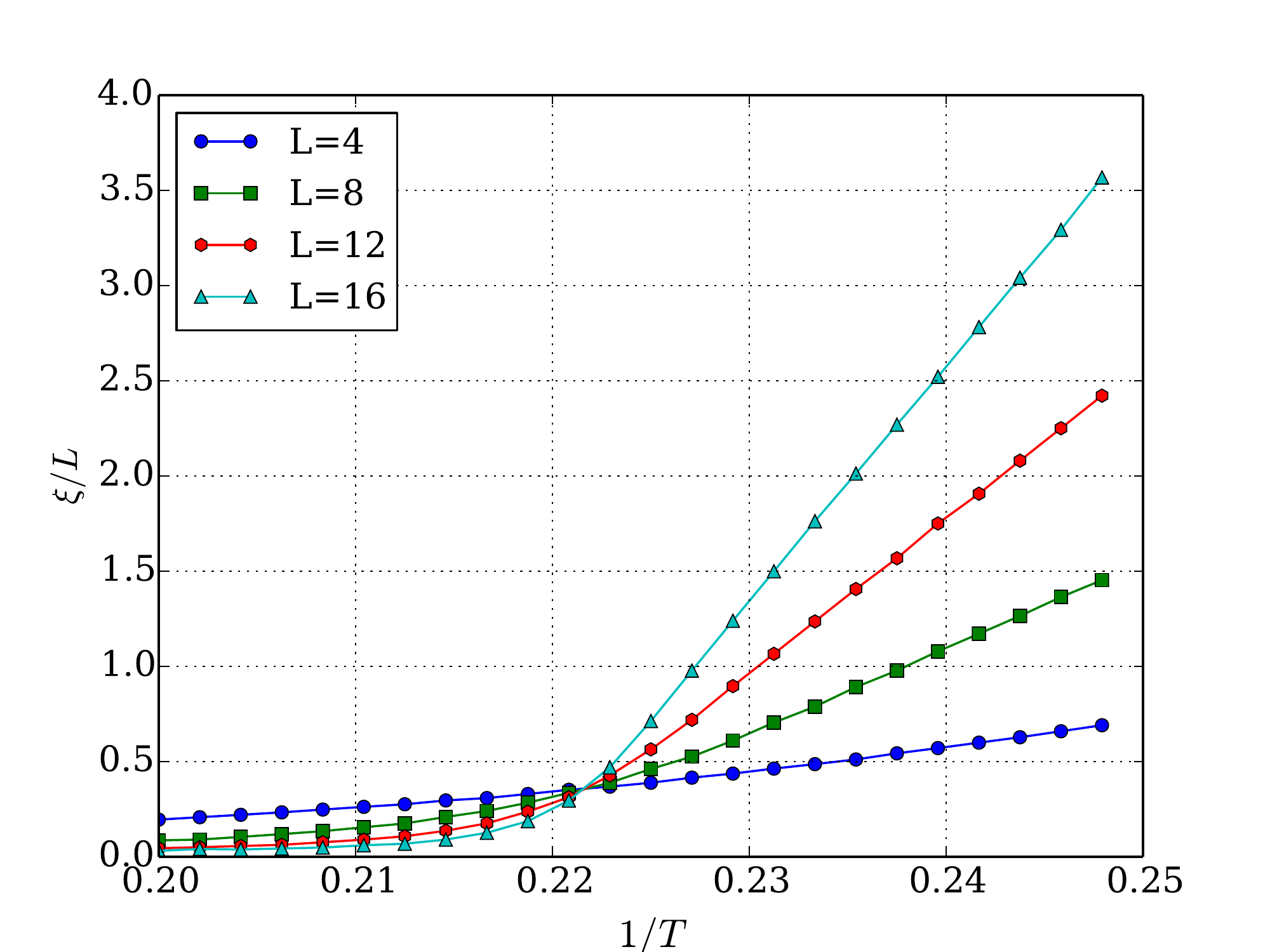}
    \caption{Correlation length vs.\ inverse temperature for the Ising model. The lines from different system sizes cross
    close to $1/T=0.2217$, which is in agreement with the published result for the critical temperature.~\cite{PhysRevB.44.5081}}
    \label{fig:corr-ising}
    \end{figure}

\begin{figure}[ht!]
    \centering
    \includegraphics[width=0.5\textwidth]{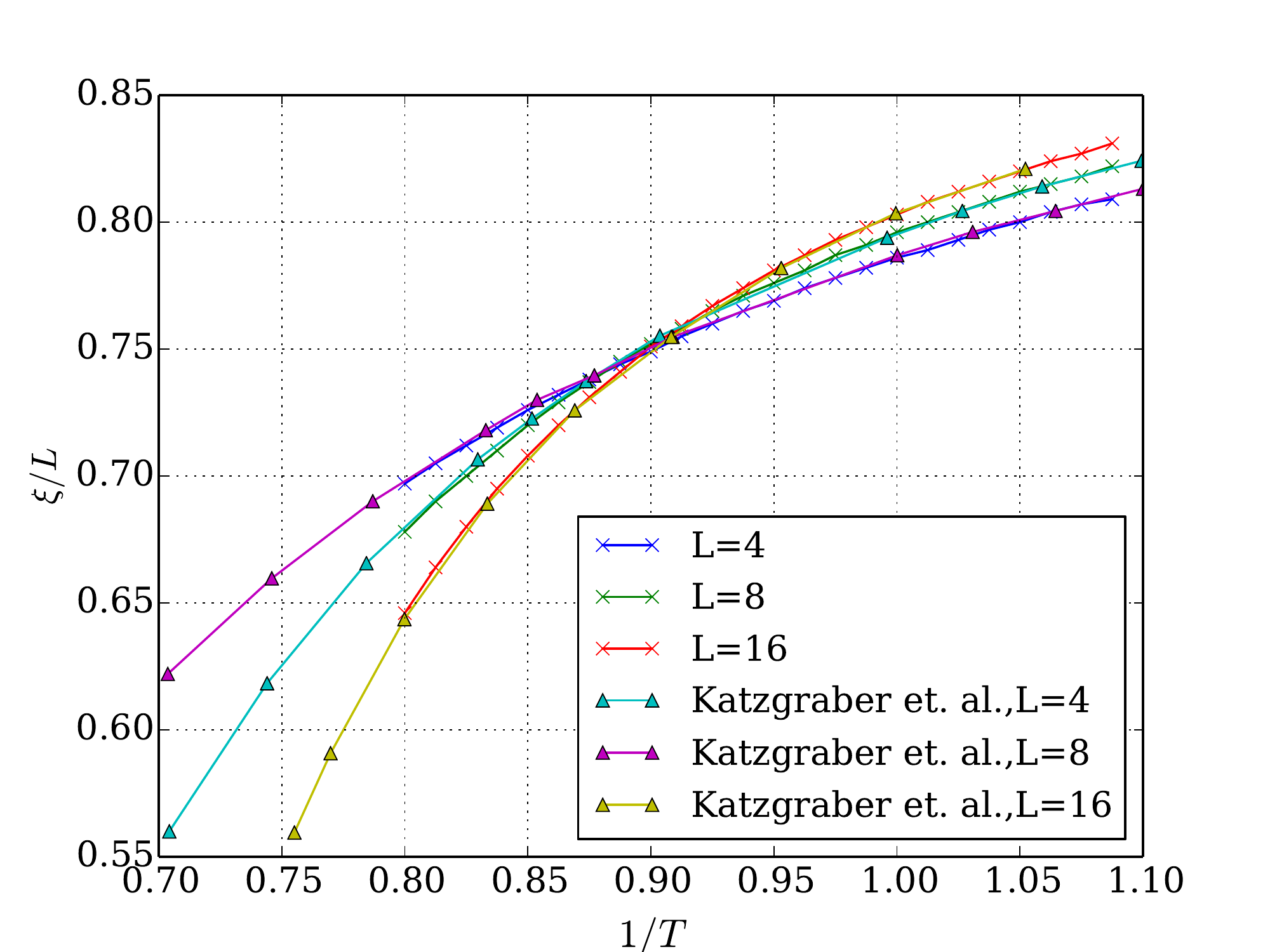}
    \caption{Binder Ratio for the 3D Edwards-Anderson model. The data generated by our GPU code is compared 
    with the data extracted from the paper by Katzgraber {\it et al.} \cite{Katzgraber-Korner-Young-2006}}
    \label{fig:binder-ea}
    \end{figure}

\begin{figure}[ht!]
    \centering
    \includegraphics[width=0.5\textwidth]{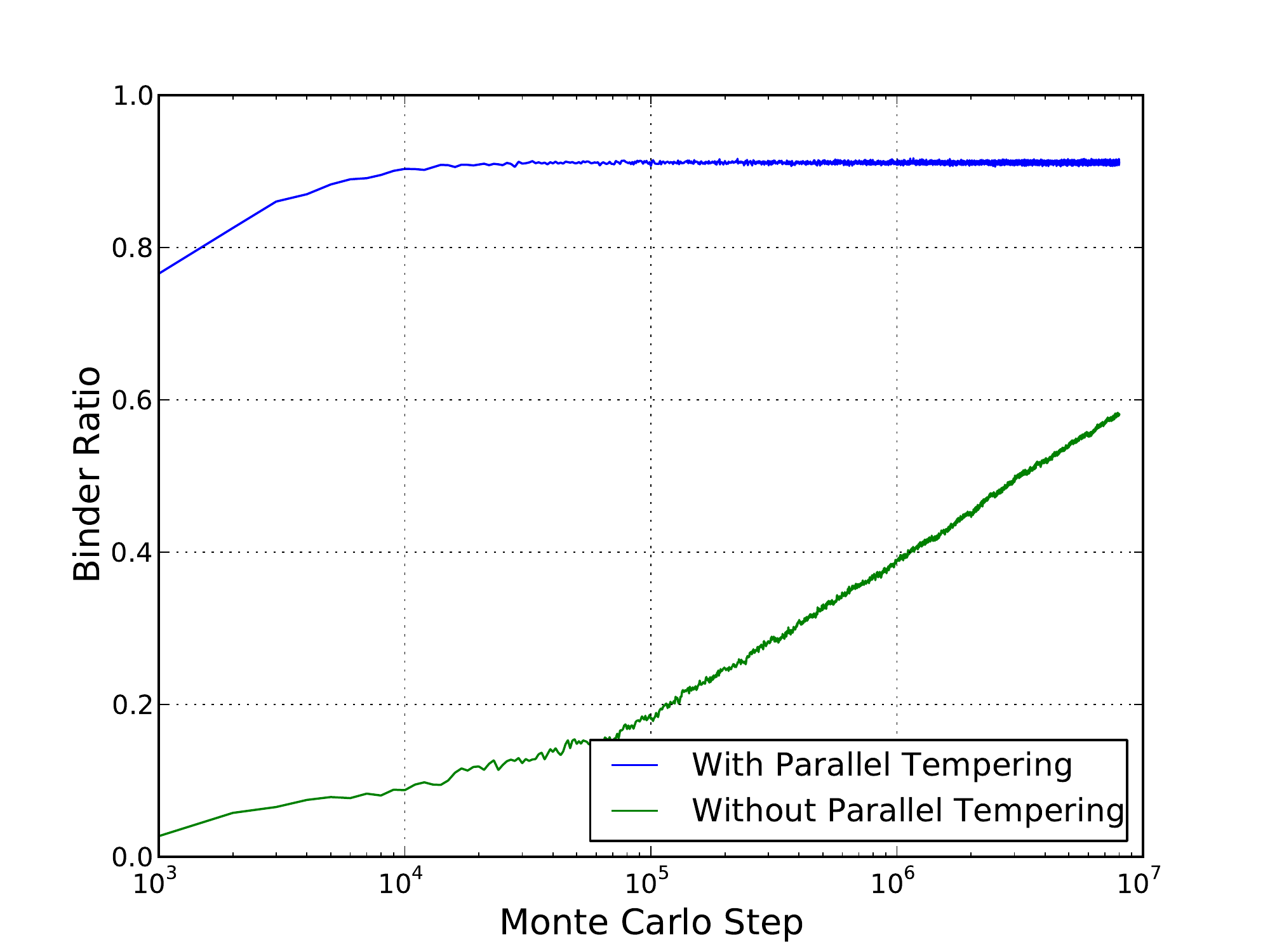}
    \caption{The convergence of the Binder ratio vs.\ number of Monte Carlo steps for the Edwards-Anderson
      model in a system with $8^3$ sites, 
    with and without parallel tempering for  $1/T=2.0$.  Parallel tempering dramatically improves 
    the convergence to equilibrium.}
    \label{fig:qvst}
    \end{figure}

We test the code by simulating  both the simple ferromagnetic
Ising and the EA spin glass models.  In Figure
\ref{fig:energy-ising}, our results from the GPU code are found to be
consistent with the results from our CPU code for the ferromagnetic
Ising model, at various external magnetic fields. We also compare the
results with and without parallel tempering as a check to determine
whether the parallel tempering swap is performed correctly. We find
that the results with and without the parallel tempering swap are
consistent with each other. In Figure \ref{fig:corr-ising} we plot
the correlation length for the ferromagnetic Ising model in three
dimensions; here, the crossing point for the correlation length coincides
with the known critical temperature for the ferromagnetic
ordering. \cite{PhysRevB.44.5081}
For the EA model we calculate the Binder ratio
of the system at zero external magnetic field as shown in Figure
\ref{fig:binder-ea}. The results match reasonably well with the
published data. \cite{Katzgraber-Korner-Young-2006} Figure \ref{fig:qvst} demonstrates the effectiveness
of parallel tempering for the EA spin glass. The
parallel-tempering simulation reaches equilibrium after $10^5$ Monte
Carlo sweeps, while without parallel tempering, the system did not
reach equilibrium even with 100 times more iterations. This further
supports that we have implemented the parallel tempering swapping
correctly.

\section{Conclusion and Future Works}

We design and implement a CUDA code for simulating the random frustrated
three-dimensional Edwards-Anderson Ising model on GPUs. 
For small to intermediate system sizes, our code runs faster than other
GPU implementations, and its speed is close to that of the specially built FPGA
computer. We note a very recent	preprint has reported an improvement in FPGA system. \cite{Janus2-2013}
Our performance tuning strategies include constructing three
levels (tasks, threads, bits) of parallel workloads for GPU;
optimizing the memory access via a proper data layout and tiling; 
speeding up the computation by translating time consuming floating point
operations to integer point operations and table look-ups; and finally,
vectorizing bit computations with our binary format, the Compact Asynchronous Multispin coding.

Our program can be extended for other models such as the Potts models and models with different random coupling distributions.
The structure of our code may adapt well to upcoming GPUs and future massive parallel accelerators.

\section*{Acknowledgments}

This work is sponsored by the NSF EPSCoR LA-SiGMA project under award number EPS-1003897. 
Portions of this research were conducted with high performance computational resources provided by 
Louisiana State University (http://www.hpc.lsu.edu).  Part of this work was done on the Oakley 
system at the Ohio Supercomputer Center. We thank Helmut Katzgraber and Karen Tomko for useful discussions. 
We thank the following collaborators: Bhupender Thakur, 
Ariane Papke, Sean Hall and Cade Thomasson. We thank Samuel Kellar for his careful reading of the manuscript.






\bibliographystyle{model1-num-names}
\bibliography{ising}







\end{document}